\documentclass{aa}  

\usepackage{graphicx}
\usepackage{txfonts}
\usepackage{longtable}
\usepackage{lscape}
\usepackage[font=normalsize]{caption}
\usepackage{subfig}

\begin{document}

   \title{Discovery of a 26.2~day period in the long-term X-ray light curve of SXP~1323: a very short orbital period for a long spin period pulsar}
   \titlerunning{Discovery of a 26.2~day period in the long-term X-ray light curve of SXP~1323}


   \author{S. Carpano
          \inst{1}
          \and
          F. Haberl\inst{1}
          \and
          R. Sturm\inst{1}
          }

   \institute{Max-Planck-Institut f\"{u}r extraterrestrische Physik, Giessenbachstrasse 1, 85748 Garching, Germany \\
              \email{scarpano@mpe.mpg.de}
             }

   \date{11 April 2017}

 
  \abstract
   {About 120 Be/X-ray binaries (BeXBs) are known in the Small Magellanic Cloud (SMC); about half of them are pulsating with periods from a few to hundreds of seconds. SXP~1323 is one of the longest-period pulsars known in this galaxy.}
  {SXP~1323 is in the field of view of a large set of calibration observations that we analyse systematically, focusing on the time analysis, in search of periodic signals.}
   {We analyse all available X-ray observations of SXP~1323 from Suzaku, XMM-Newton, and Chandra,  in the time range from 1999 to the end of 2016. We perform a Lomb-Scargle periodogram search in the band 2.5-10\,keV on all observations to  detect the neutron star spin period and constrain its long-term evolution. We also perform an orbital period search on the long-term light curve, merging all datasets.}
   {We report the discovery of a 26.188$\pm$0.045~d period analysing data from {Suzaku, XMM-Newton, and Chandra}, which confirms the optical period derived from the Optical Gravitational Lensing Experiment (OGLE) data. If this corresponds to the orbital period, this would be very short with respect to what is expected from the spin/orbital period relationship. We furthermore report on the spin period evolution in the last years. The source is spinning-up with an average rate of $\lvert\dot{P}/P\lvert$ of 0.018 yr$^{-1}$, decreasing from  $\sim$1340 to $\sim$1100 s, in the period from 2006 to the end of 2016, which is also extreme with respect to the other Be/X-ray pulsars. From 2010 to the end of 2014, the pulse period is not clearly detectable, although the source was still bright.}
   {SXP~1323 is a peculiar BeXB due to its long pulse period, rapid spin-up for several years, and short orbital period. A continuous monitoring of the source in the next years is necessary to establish the long-term behaviour of the spin period.}


   \maketitle
%

\section{Introduction}
Be/X-ray binaries (BeXBs) belonging to the class of High-Mass X-ray binaries (HMXBs) are composed of a Be star and a compact object, generally a neutron star. In those systems, it is believed that mass transfer occurs from the equatorial decretion disc around the donor star, onto the compact object during the periastron passage of the neutron star in an eccentric orbit, either via an accretion disk or via wind capture.
In this class of object, the pulse period is generally well correlated to the orbital period as reported initially by \cite{Corbet1984, Corbet1986}, although with large scatter. The second group of HMXBs are the supergiant X-ray binaries, where the massive companion is an evolved star in which matter is transferred to the compact object via a strong wind.

A large number of BeXBs have been reported so far in the Small Magellanic Cloud (SMC) \citep{Coe2015, Haberl2016}. \cite{Galache2008} monitored 41  BeXB systems in the SMC over nine years using RXTE  Proportional Counter Array (PCA) data in search of orbital modulations. They confirmed  and refined ten known orbital ephemerides and determined ten new ones. More recently, \cite{Klus2014} reported the long-term average spin period of 42 BeXB systems, using RXTE data, claiming they all contain a neutron star accreting via a disc rather than a wind.

SXP~1323 is a pulsar discovered by \cite{Haberl2005}, and it shows one of the longest pulse periods known in the SMC.
\cite{Schmidtke2006, Schmidtke2006b} reported the discovery of several strong periodic signals in the optical light curve of SXP~1323, using Optical Gravitational Lensing Experiment (OGLE) data. The optical light is coming from the Be star itself, the decretion disk and transient accretion disk around the neutron star. Three strong peaks were discovered at periods of 0.41\,d, 0.88\,d, and 26.16\,d, all showing approximatively sinusoidal light curves. The first two are believed to come from non-radial pulsations of the Be star.  Later, \cite{Bird2012}, analysed the OGLE light curves of 49 SMC BeXBs, and confirmed  the period at 26.17\,d. Authors of both papers conclude that the short periods and sinusoidal shapes are not characteristic of an orbital modulation of a decretion disc, and explain the 26.2\,d period with the aliasing of non-radial pulsations at a period of 0.96\,d.

\section{Observations and data reduction}
\label{sec:observ}

SXP~1323 is very close to the bright supernova remnant (SNR) 1E~0102-72.3, which is often observed for calibration purposes \citep[see e.g.][]{Plucinsky2008, Plucinsky2016}.  A large number of observations of the source are therefore available from several X-ray observatories.

\subsection{Suzaku observations}
\label{sec:suzobs}

 \begin{figure}
   \centering
    \resizebox{\hsize}{!}{\includegraphics{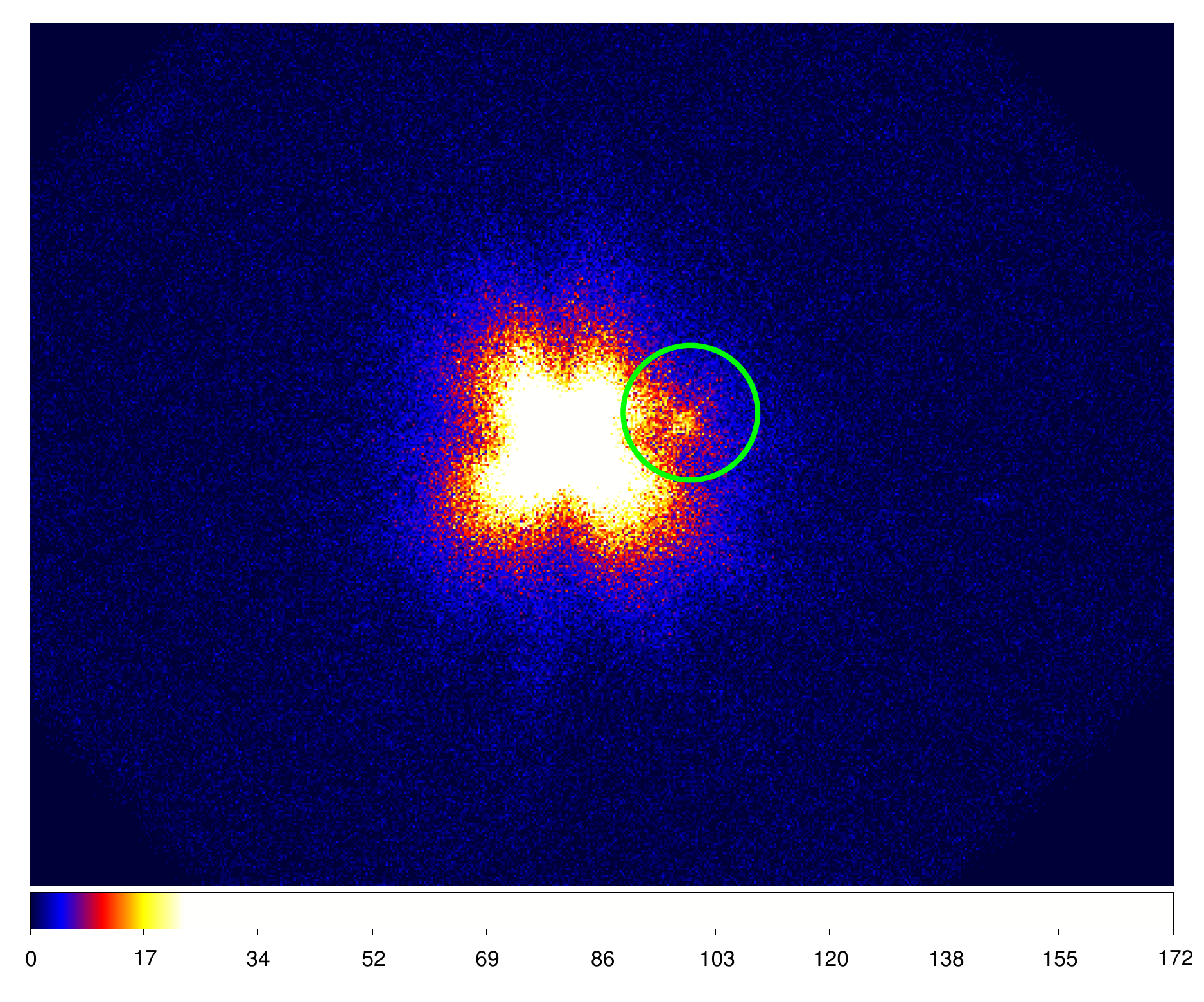}}
    \resizebox{\hsize}{!}{\includegraphics{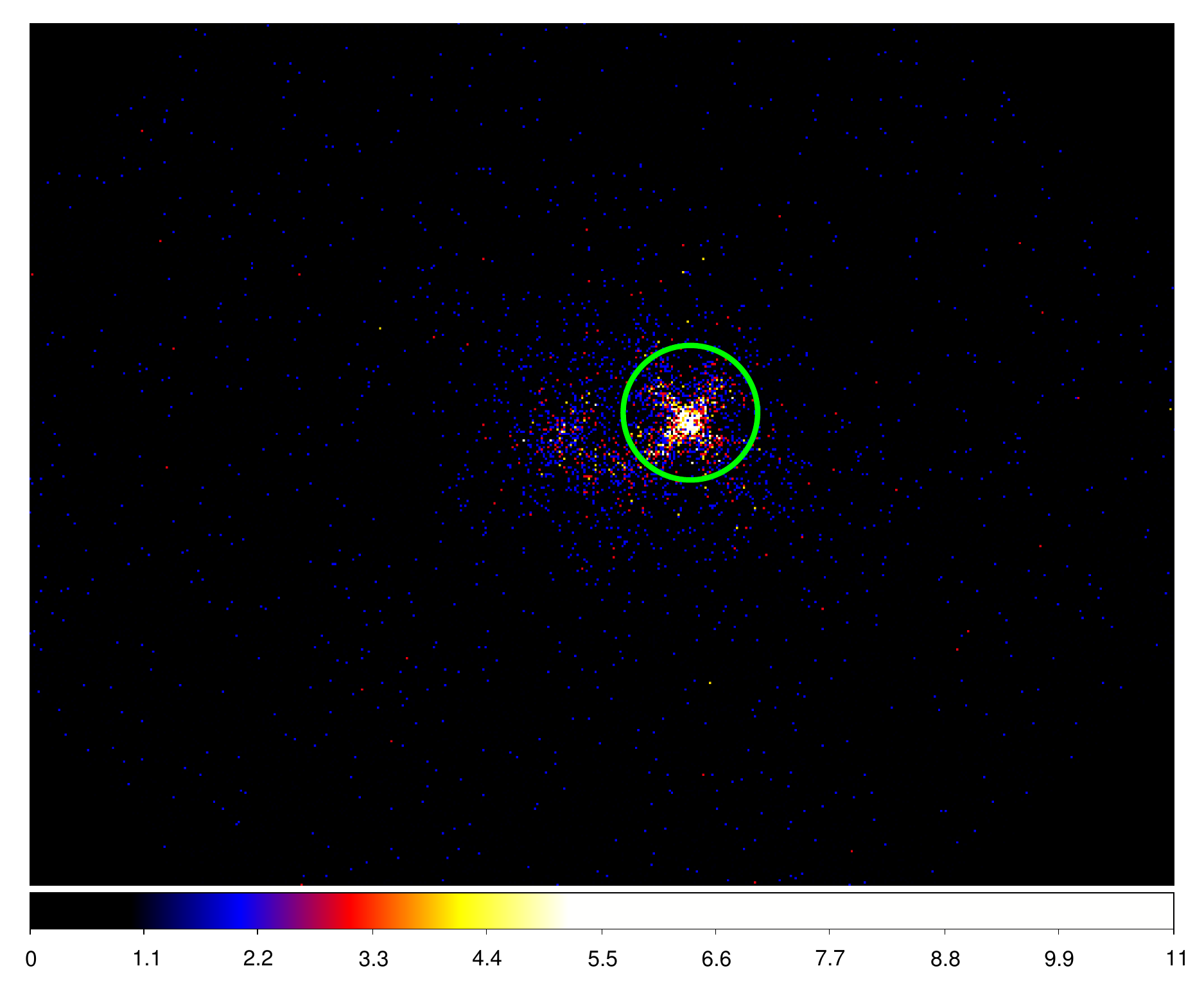}}
   \caption{Suzaku images (in counts) of the SNR 1E~0102-72.3 (obsID: 101005050), in the full energy band (top) where the pulsar and the SNR are confused, and in the 2.5-10\,keV band (bottom) where the contribution of the SNR is negligible.}
              \label{fig_imsuz}
 \end{figure}

1E~0102-72.3 has been observed 75 times with Suzaku \citep{Mitsuda2007}, between August 2005 and April 2015 with the X-Ray Imaging Spectrometer (XIS) instrument \citep{Koyama2007}. We performed a period search on 70 XIS datasets  having a total exposure time longer than 15\,ks and we use all events from cleaned event files read out in 3x3 and 5x5 modes.
A summary of all datasets used in this analysis is given in Table~\ref{tab:suz_obs}. The first, second, and third columns give the observation ID, the instrument name (XIS0, XIS1, XIS2 and XIS3), and the readout mode (3x3 or 5x5 pixels read around the centre of each event), respectively. The following three columns give the date of the start of the observations, the time elapsed from the beginning to the end of the observation (in seconds) and the exposure time (in seconds), taken from the \texttt{ONTIME} keyword.
In the full energy band, the light of the standard candle SNR~1E~0102-72.3 and the pulsar are confused. However, if events are filtered, with energies >2.5\,keV, the contribution of the SNR is negligible, as shown in Fig.~\ref{fig_imsuz}.

\subsection{XMM-Newton observations}
Currently, 48 XMM-Newton \citep{Jansen2001} observations are available for SNR 1E~0102-72.3. For two of them no European Photon Imaging Camera (EPIC) data are available, leaving 46 datasets with exposures from $\sim$14 to 69\,ks performed between April 2000 and December 2016. Only exposures recorded in imaging mode and with the source in the field of view are analysed. Table~\ref{tab:xmm_obs} reports a summary of all observations used in this work.  

For the event filtering, single to double and single to quadruple events are used for pn \citep{Strueder2001} and Metal-Oxide-Silicon (MOS) \citep{Turner2001} cameras respectively, all with FLAG=0.

\subsection{Chandra observations}
\label{sec:chanobs}

The pulsar is visible in 201 observations performed with Chandra's Advanced CCD Imaging Spectrometer (ACIS) instrument \citep{Garmire2003}, on the SNR. Those data were recorded from August 1999 to March 2016 and the datasets used in this paper are listed in Table~\ref{tab:chan_obs}. Events from observations  performed on the same day or on two consecutive days were merged together, leaving  66 independent datasets (see column 1 of Table~\ref{tab:chan_obs}). Columns 2 and 3 give the observation ID, the operating ACIS CCDs (0-3 are I0-I3 and 4-9 are S0-S5). From 2008, the source is not in the field-of-view for most observations since only one detector was used to observe 1E~0102-72.3.

\section{Analyses and results}

The source extraction region is centred on coordinates RA, DEC= 01:03:37.8,  -72:01:33, extracted from the Two Micron All Sky Survey (2MASS) catalogue \citep{Cutri2003}, with a radius of  62.5\,\arcsec\, for Suzaku and 30\,\arcsec\ for  XMM-Newton. For Chandra the radius was 7.4\,\arcsec, 12.3\,\arcsec, 24.6\,\arcsec\, depending on the off-axis angle of the source.  For the background region, for XMM-Newton and Chandra, we use an annulus centred on the source, both with an inner and outer radius of, for XMM-Newton, r$_\textrm{in}$,  r$_\textrm{out}$=25\,\arcsec, 40\,\arcsec\ and for Chandra, depending on the position of the source in the field of view: r$_\textrm{in}$,  r$_\textrm{out}$=7.4\,\arcsec, 14.8\,\arcsec, r$_\textrm{in}$,  r$_\textrm{out}$=12.3\,\arcsec, 24.6\,\arcsec\, or r$_\textrm{in}$,  r$_\textrm{out}$=24.6\,\arcsec, 39.6\,\arcsec. On the other hand, in the case of  Suzaku, the background region was not centred around the pulsar coordinates because of a potential contamination from the nearby SNR due to the large extraction region. Instead it is centred for most of the observations around coordinates RA, DEC= 01:02:56, -72:00:57, with a radius of  62.5\,\arcsec\, (another region is chosen when this one is out of the field of view).

\subsection{Pulse period search}
\label{sec:pulse}

\begin{figure}
   \centering
   \resizebox{\hsize}{!}{\includegraphics{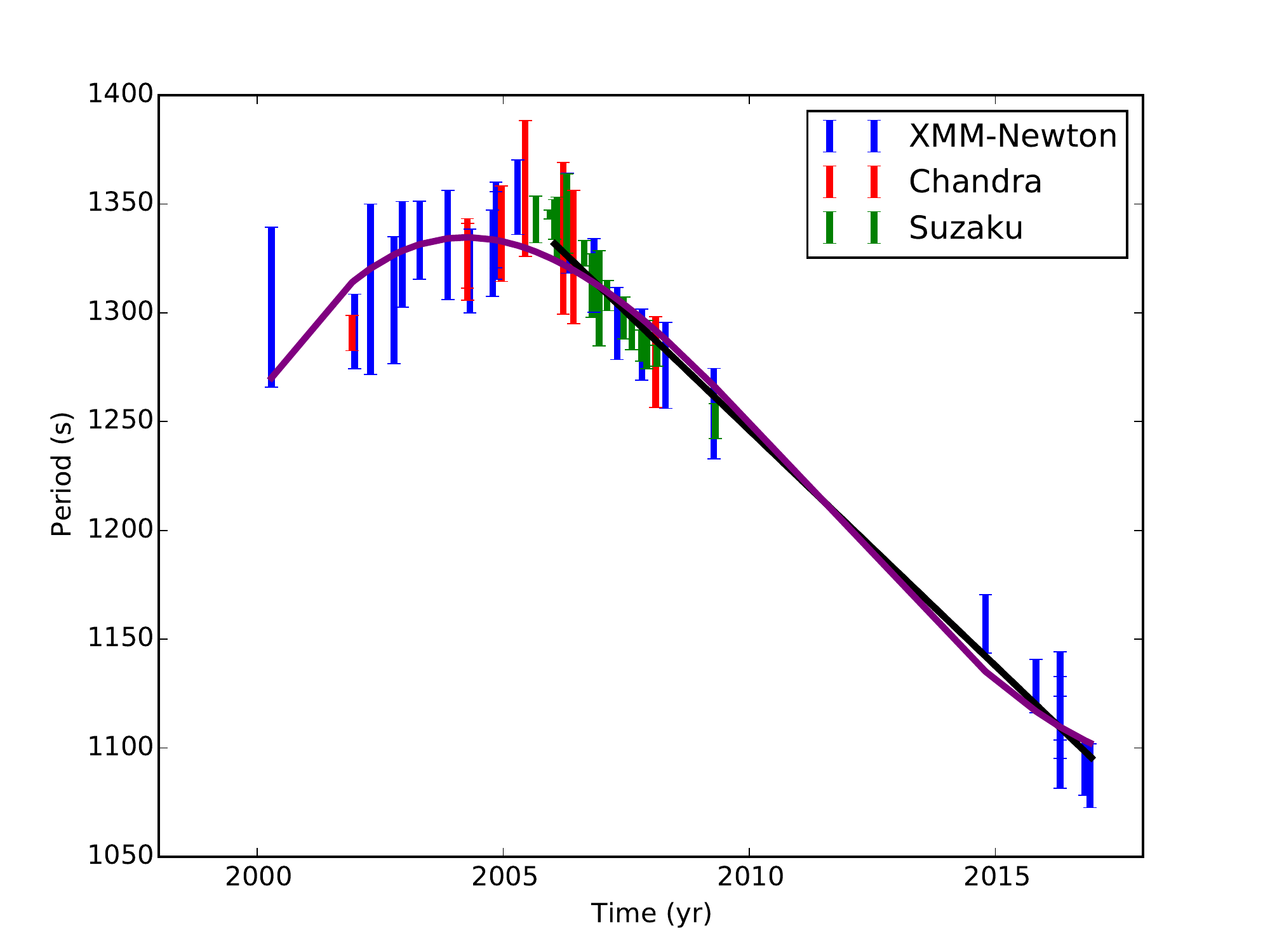}}
   \caption{Long-term spin period evolution observed with Suzaku  (green), XMM-Newton (blue) and Chandra (red) data. The dotted black line is the result of a linear fit performed on data from 2006 to 2017.}
              \label{fig_pertime}      
\end{figure}

The pulse period search was performed using Lomb-Scargle periodogram analysis \citep{Lomb1976, Scargle1982}, in the energy band 2.5--10\,keV where the signal-to-noise ratio for the emission of the pulsar is maximised, and in the period range 900--1600\,s. The light curves are  barycenter-corrected, background-subtracted and binned to 50\,s. 
In the case of XMM-Newton and Suzaku, events recorded by the different instruments are merged together. However, when one or more instrument is not operating for some time interval discontinuities appear in the light curves. We correct for those effects by multiplying the portions of the light curves not covered by all cameras, by some factor, assuming that, in the 2.5--10.0\,keV band, the effective area of the XMM-Newton pn camera is about 3.2 times larger than the MOS ones, and that the effective area of Suzaku's front illuminated cameras (XIS0, XIS2 and XIS3) is about 1.1 times larger than the back illuminated one (XIS1). The uncertainties on the period are measured by fitting a Gaussian function with the \texttt{Python} module \texttt{astropy.modeling.Gaussian1D}. The error bars correspond to the 1-$\sigma$ standard deviation of the Gaussian.

To calculate the confidence levels, since the noise is not purely white photon noise, but also contains some red/correlated component, we proceed in the following way:
\begin{enumerate}
\item We divide light curves into blocks of maximum 1000\,s for Chandra and XMM-Newton, and of 4000\,s for Suzaku (for which observations are lasting much longer), ensuring that at least ten blocks are present per light curve.
\item We shuffle the different blocks randomly, perform a Lomb-Scargle period search and determine the maximal power of the corresponding periodogram.
\item We repeat  the 2nd step 1000 times and, by collecting the maxima of all periodograms, we derive the confidence levels.
\end{enumerate}

Results of the spin period search are shown in Fig.~\ref{fig_pertime}, for Suzaku (green), XMM-Newton (blue), and Chandra (red) data. Only periods with significance above 99$\%$ are shown. 
From $~$2002 to 2006 the pulse period seems to have increased slightly, although error bars are relatively large. This spin-down trend is in contradiction with the slow spin-up trend reported by \cite{Klus2014} who analysed RXTE data in roughly the same time interval. 

From 2006 to end 2016, the source is rapidly spinning up at a rate of $\dot{P}=-21.65$\,s\,yr$^{-1}$. This value was calculated by fitting a line using the \texttt{numpy.polyfit} function in \texttt{Python} (shown as a black dotted line in Fig.~\ref{fig_pertime}). Using the mean value of 1216\,s in the 2005--end 2016 time interval, the relative spin period change becomes: $\lvert\dot{P}/P\lvert$=0.0178\,s\,yr$^{-1}$\,s$^{-1}$. Alternatively, we tried to fit a 3rd degree polynomial (in purple) that takes into account the data before 2005 as well, while a 2nd degree polynomial alone would not fit the data well. From 2010 to the end  of 2014, the pulse period is only occasionally detected, but at a lower confidence level.

\subsection{The orbital period search}
\begin{figure}
   \centering
   \resizebox{\hsize}{!}{\includegraphics{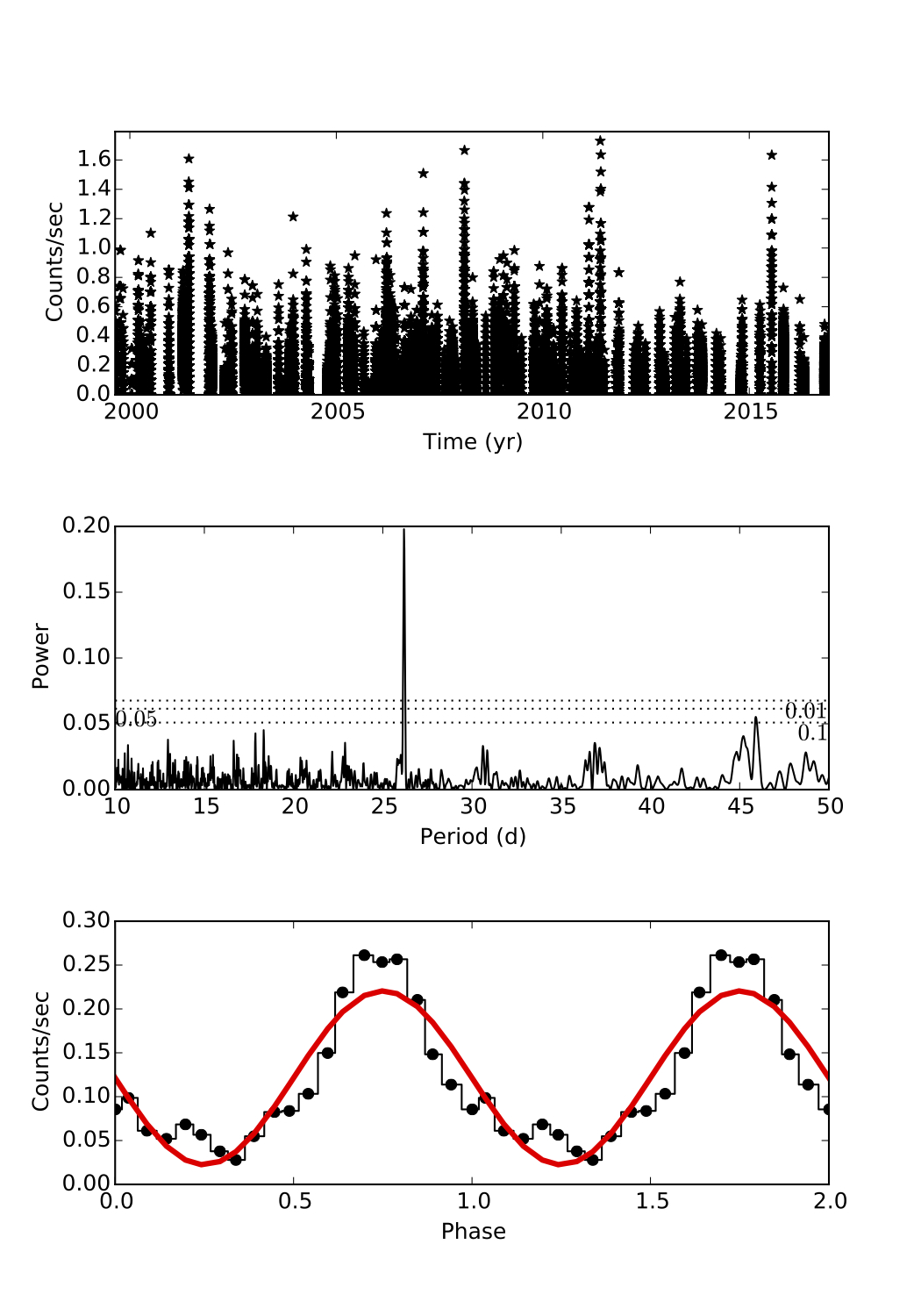}}
   \caption{Top: merging of all Suzaku, XMM-Newton, and Chandra observations as described in section~\ref{sec:observ} (arbitrary scale). Middle: Lomb-Scargle periodogram on all observations, performed over a period range of 10 to 50\,d with a peak at 26.19\,d. Bottom: folded light curve with fitted sine function.}
              \label{fig_orbit}
              
\end{figure}

The search for the orbital period was made using all available observations from Suzaku, XMM-Newton and Chandra merged together and a period search was performed in the range from 10 to 50\,d. Differences in the effective area from the various instruments and vignetting issues had to be taken into account. We therefore created a simulated spectrum for the source in every observation, using \texttt{XSPEC}  and the \texttt{fakeit} command and using the relevant response files.
The model spectrum is the one described in Section~\ref{sec:spec}, with $N_\textrm{H}$ value of 0.5$\times$10$^{22}$\,cm$^{-2}$ and  $\Gamma$=0.7. The count rate extracted in the 2.5--10\,keV band, from all spectra, is recorded and then used to scale the different light curve segments, since these represent the values expected for a constant source. We note that, for every observation, we extract and scale the expected count rates as if all instruments were always operating (pn, MOS1 and MOS2 for XMM-Newton, and XIS0, XIS1, XIS2 and XIS3 for Suzaku). We then divide, for each observation, the measured count rate by its theoretical counterpart and multiply by a single factor to get values expected for XMM-Newton (pn+MOS1+MOS2).
The results  are shown in Fig.~\ref{fig_orbit}, where the total background-subtracted and barycenter-corrected light curve is shown at the top, the Lomb-Scargle periodogram in the middle and the folded light curve with sinusoidal fit  at the bottom. A clear peak is found at a period of 26.188\,d and a 1-$\sigma$ uncertainty of 0.045\,d is determined from the width of the Gaussian fit on the peak.  Phase 0 of the sine function is at MJD=51417.1695288$\pm$N$\times$26.188$\pm$0.045, where N is an integer (maximum is at phase 0.75).
To calculate the significance levels we again use the bootstrap method but this time we randomly shuffle  the whole light curves of all observations, again 1000 times.

\subsection{Spectral changes with orbital period}
\label{sec:spec}
Because of its higher sensitivity, XMM-Newton data are analysed to search for any change in the X-ray spectra with the orbital modulation. We extracted a source and background spectrum for each observation and exposure separately using standard \texttt{SAS} tasks. We fit the data,  in the 0.3-10\,keV energy band, with \texttt{XSPEC} using a power-law emission component and two absorption components (\texttt{phabs*vphabs*power}). The first absorption component has a fixed Galactic $N_\textrm{H}$ value of 5.36$\times$10$^{20}$\,cm$^{-2}$ \citep{Dickey1990} and the second $N_\textrm{H}$ is left free, while the abundance  is fixed to 1.0 for He and 0.2 for Z>2 \citep{Russell1992}. Sometimes a faint soft excess (<1\,keV) is observed but because this component it not well constrained, we use only the power-law component to model the spectrum.

\begin{figure}
   \centering
   \resizebox{\hsize}{!}{\includegraphics{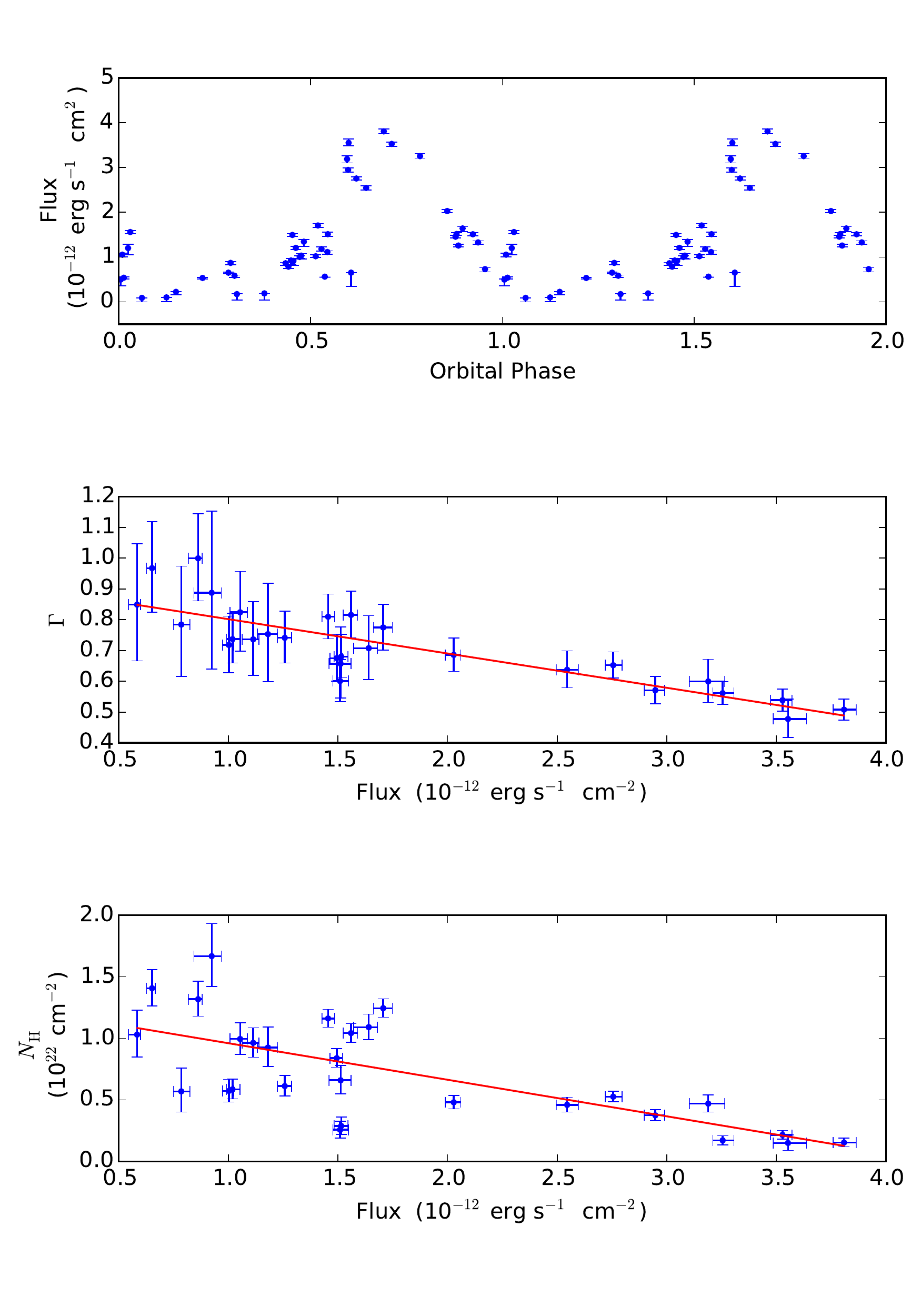}}
   \caption{0.3-10\,keV fluxes derived from spectral fits for XMM-Newton (top) as a function of orbital phase (phase 0 is the same as for the merged data). Best fitted $\Gamma$  value for the power-law component (middle) and $N_\textrm{H}$ for the absorption (bottom) as a function of flux.}
              \label{fig_spec}             
\end{figure}

The results of the spectral fits are shown in Fig.~\ref{fig_spec}. The plot at the top shows the observed 0.3-10\,keV flux as a function of the orbital phase; errors are given at 68$\%$ confidence level. From those plots we can observe a variation of  the flux with the orbital phase (phase 0 is the same as for the merged data in Fig.~\ref{fig_orbit}, that is, MJD=51417.1695288).
In the second and third plots, values of $\Gamma$ for the power-law component and  $N_\textrm{H}$ for the absorption component are shown, as a function of the flux. 

Fitting a linear relation between the parameters and the flux, gives slopes of -0.11 x 10$^{12}$ erg$^{-1}$~s~cm$^2$ for the $\Gamma$ and -0.30  x 10$^{34}$ erg$^{-1}$~s  for the $N_{\textrm{H}}$ although there is a larger scatter in the latter plot, indicating that there is an anti-correlation between the power law index and the flux (the source getting harder  at higher fluxes) and between the absorption and the flux (less absorption at higher fluxes), and hence with the orbital period.

\section{Discussion}
From the analysis of XMM-Newton,  Suzaku, and  Chandra data it appears that SXP~1323 is a quite atypical BeXB. First, it has a long pulse period, with a very high derivative, and as described in Sec~\ref{sec:pulse}, from 2006 to the end of 2016, the neutron star is spinning up at a very rapid average rate: $\lvert\dot{P}/P\lvert$=0.018\,s\,yr$^{-1}$\,s$^{-1}$. This is much higher than for any other SMC BeXB reported by \cite{Klus2014}  in their Table~4. From 2010 to the end of 2014, the pulse period is no longer clearly detected.  Such rapid spin-up and temporary apparent pulsation disappearance for several years has also been reported by \cite{Townsend2013} on SXP~91.1 using RXTE data. In their Fig.~6, they report a spin period changing linearly from 91.1\,s to 85.4\,s in approximatively 13.5\,yrs. A linear fit provided a spin-up rate of $\dot{P}=1.442 \times10^{-8}$\,s s$^{-1}$. Dividing by the average period (88.45\,s), and changing units leads to $\lvert\dot{P}/P\lvert$=0.005\,s\,yr$^{-1}$\,s$^{-1}$, which is still smaller than what we derived for SXP~1323. From MJD $\sim$52500 to $\sim$55200, that is, for about 8\,yrs, no pulse period could be found for SXP~91.1. The authors suggest that the source is continuously accreting matter, and that it is unlikely that the source stopped to pulse for many years. Instead there might be nearby absorbing material preventing the pulsed X-ray emission from reaching the observer, or there might be a change in the geometry. This could easily be concluded for SXP~1323 as well. 

A second peculiarity of the pulsar reported here, SXP~1323, is the very short 26.2\,d period detected in X-rays and optical wavelengths compared to the long pulse period. While the 26.2\,d optical period has been explained so far by aliases of non-radial pulsations of the Be star, this is no longer possible due to the detection of the same period in X-rays. Although we cannot exclude other possibilities for the X-ray variability, we don't know of any plausible explanation which is not related to the orbital period. Such a short orbital period does not follow  the spin-orbit period relationship reported by \cite{Corbet1984, Corbet1986}, although there is a large scatter of the observed systems for this relation. From Eq.~1, in \cite{Corbet1986}, a pulse period of 1323\,s would lead to an orbital period of 364\,d for circular orbit, and higher for an eccentric orbit.  This relationship assumes that the neutron star is in a state of quasi-equilibrium in which the Alfv\'{e}n and corotation radii are equal on average. Other exceptions to this rule are observed for BeXBs in the SMC, although these are less extreme cases than SXP~1323. Some examples with $P_\textrm{spin}$ >100\,s and $P_\textrm{orb}$<50\,d are SXP~264 \citep{Haberl2004} with an orbital period of 49.1\,d  \citep{Schmidtke2005}; SXP~325 \citep{Haberl2008} with an orbital period of 45.995\,d \citep{Coe2008}; SXP~214 \citep{Coe2011} with an orbital period of 29.91\,d \citep{Schmidtke2013}; and SXP~101 \citep{Yokogawa2000} with an orbital period  of 21.94 \citep{McGowan2007}. All orbital periods were derived using optical data from OGLE and in some cases from the Massive Astrophysical Compact Halo Object (MACHO) project as well.
In our Galaxy, some other comparable systems are: 1A~1118-616, with a pulse period of 405\,s and an orbital period of 24.0\,d \citep{Staubert2011} determined using RXTE data, and, SAX J2103.5+4545 with a pulse period of 358.6\,s, which is the BeXB with the shortest known orbital period of 12.7\,d \citep{Reig2004} determined using RXTE data as well. They occupy the region of the wind-fed supergiant binaries in the $P_\textrm{spin} - P_\textrm{orb}$ diagram. \cite{Reig2010} suggest that the X-rays observed during the quiescence state of SAX J2103.5+4545 are more likely the result of wind accretion. The dominant accretion wind is likely coming from a more stable equatorial low-velocity high-density wind from the Be star, since, in contrast to wind-fed systems with supergiants, which show erratic and flaring X-ray variability, the X-ray flux in quiescence is stable. In this scenario, the neutron star orbit is coplanar to the plane of the circumstellar disk.

Such a model could be applied to SXP~1323 as well, to explain on one hand the presence of a short orbital period and on the another hand the sinusoidal shapes of the folded X-ray and the optical light curves, as shown in this work by \cite{Schmidtke2006} and \cite{Bird2012}. This rather sinusoidal shape would mean that the system is not undergoing regular outbursts as commonly observed in BeXBs, but the neutron star seems to accrete continuously during its orbit around the Be star companion. Accretion from an equatorial low-velocity high-density wind could provide an explanation. Some of the SMC BeXBs with known X-ray orbital periods presented by \cite{Galache2008} also have sinusoidal shape, similar to, for example, SXP~138 ($P_\textrm{orb}$=103.6\,d) and SXP~280 ($P_\textrm{orb}$=64.8\,d) and could also be explained by this model.

\section{Conclusions}
From this work, it appears that SXP~1323 is a peculiar BeXB due to its long pulse period, rapid spin-up for several years, and short orbital period. Thanks to the nearby SNR 1E~0102-72.3, which is used as a calibration target for most CCD detectors, a large number of observations are already available for this source allowing a study of the long-term behaviour of the pulse period and the detection of long periodicities in X-rays. 
A continuous monitoring of the source in the future is necessary to determine how the pulse period will evolve in the near future.

\begin{acknowledgements}     
The scientific results reported in this article are based on data obtained from the Chandra Data Archive, on observations obtained with XMM-Newton, an ESA science mission with instruments and contributions directly funded by ESA Member States and the USA (NASA), and on data obtained from the Suzaku satellite, a collaborative mission between the space agencies of Japan (JAXA) and the USA (NASA). 
 
\end{acknowledgements}

\bibliographystyle{aa}
  \bibliography{article} 
  
\begin{appendix}
\section{Summary of the XMM-Newton, Chandra and  Suzaku observations}  
\captionsetup[table]{labelformat=empty}
\captionsetup[subfloat]{labelformat=empty}
\onecolumn

\begin{table}
\caption{Table~A.1: Summary of the Suzaku observations used in this work (see text for more details).}
\begin{scriptsize}
\subfloat{
\begin{tabular}[t]{|l l l l l l|}
\hline
Obs ID & Instr & Mode & Date start & Telapse & Ontime \\ 
\hline 
100014010 & XIS0 & 3x3 & 2005-08-31T01:42:20 & 43384 & 18054 \\ 
& XIS0 & 5x5 & & 35163 & 6280 \\ 
& XIS1 & 3x3 & & 43384 & 18054 \\ 
& XIS1 & 5x5 & & 35163 & 6280 \\ 
& XIS2 & 3x3 & & 43384 & 18054 \\ 
& XIS2 & 5x5 & & 35163 & 6280 \\ 
& XIS3 & 3x3 & & 43384 & 18054 \\ 
& XIS3 & 5x5 & & 35163 & 6280 \\ 
100044010 & XIS0 & 3x3 & 2005-12-16T01:43:08 & 289212 & 46991 \\ 
& XIS0 & 5x5 & & 195874 & 12728 \\ 
& XIS1 & 3x3 & & 289212 & 93308 \\ 
& XIS1 & 5x5 & & 195874 & 12757 \\ 
& XIS2 & 3x3 & & 289212 & 46991 \\ 
& XIS2 & 5x5 & & 195874 & 12732 \\ 
& XIS3 & 3x3 & & 289212 & 46991 \\ 
& XIS3 & 5x5 & & 195874 & 12730 \\ 
100044020 & XIS0 & 2x2 & 2006-01-17T23:22:57 & 12142 & 6885 \\ 
& XIS0 & 3x3 & & 195295 & 15657 \\ 
& XIS0 & 5x5 & & 198072 & 18726 \\ 
& XIS1 & 2x2 & & 12142 & 6893 \\ 
& XIS1 & 3x3 & & 86960 & 13623 \\ 
& XIS1 & 3x3 & & 2169 & 2169 \\ 
& XIS1 & 5x5 & & 198072 & 19167 \\ 
& XIS2 & 2x2 & & 12142 & 6885 \\ 
& XIS2 & 3x3 & & 195295 & 15681 \\ 
& XIS2 & 5x5 & & 198072 & 18695 \\ 
& XIS3 & 2x2 & & 12142 & 6893 \\ 
& XIS3 & 3x3 & & 86960 & 13548 \\ 
& XIS3 & 3x3 & & 2169 & 2169 \\ 
& XIS3 & 5x5 & & 198072 & 18751 \\ 
100044030 & XIS0 & 3x3 & 2006-02-02T20:19:20 & 19746 & 11324 \\ 
& XIS0 & 5x5 & & 29241 & 9512 \\ 
& XIS1 & 3x3 & & 19754 & 11332 \\ 
& XIS1 & 5x5 & & 29241 & 9512 \\ 
& XIS2 & 3x3 & & 19754 & 11332 \\ 
& XIS2 & 5x5 & & 29241 & 9512 \\ 
& XIS3 & 3x3 & & 19738 & 11316 \\ 
& XIS3 & 5x5 & & 29241 & 9512 \\ 
101005010 & XIS0 & 3x3 & 2006-04-16T09:42:04 & 43672 & 21340 \\ 
& XIS1 & 3x3 & & 43672 & 21324 \\ 
& XIS2 & 3x3 & & 43672 & 21332 \\ 
& XIS3 & 3x3 & & 43672 & 21340 \\ 
101005020 & XIS0 & 3x3 & 2006-05-21T17:16:07 & 36774 & 17238 \\ 
& XIS1 & 3x3 & & 36774 & 18194 \\ 
& XIS2 & 3x3 & & 36774 & 17238 \\ 
& XIS3 & 3x3 & & 36774 & 18194 \\ 
101005030 & XIS0 & 3x3 & 2006-06-26T20:47:26 & 46952 & 17576 \\ 
& XIS0 & 5x5 & & 18818 & 4140 \\ 
& XIS1 & 3x3 & & 26221 & 15040 \\ 
& XIS1 & 3x3 & & 6789 & 2512 \\ 
& XIS1 & 5x5 & & 18818 & 3956 \\ 
& XIS2 & 3x3 & & 46952 & 17576 \\ 
& XIS2 & 5x5 & & 18818 & 4140 \\ 
& XIS3 & 3x3 & & 26213 & 15032 \\ 
& XIS3 & 3x3 & & 6789 & 2512 \\ 
& XIS3 & 5x5 & & 18818 & 3956 \\ 
101005040 & XIS0 & 3x3 & 2006-07-17T06:22:33 & 54584 & 12813 \\ 
& XIS0 & 3x3 & & 1391 & 1391 \\ 
& XIS0 & 5x5 & & 25535 & 7775 \\ 
& XIS1 & 3x3 & & 54584 & 14228 \\ 
& XIS1 & 5x5 & & 25535 & 7876 \\ 
& XIS2 & 3x3 & & 54584 & 12813 \\ 
& XIS2 & 3x3 & & 1391 & 1391 \\ 
& XIS2 & 5x5 & & 25535 & 7775 \\ 
& XIS3 & 3x3 & & 54576 & 14220 \\ 
& XIS3 & 5x5 & & 25535 & 7876 \\ 
101005050 & XIS0 & 3x3 & 2006-08-25T04:55:35 & 70406 & 33157 \\ 
& XIS0 & 5x5 & & 24602 & 7263 \\ 
& XIS0 & 5x5 & & 3648 & 1526 \\ 
& XIS1 & 3x3 & & 98024 & 57716 \\ 
& XIS1 & 3x3 & & 1846 & 1846 \\ 
& XIS1 & 5x5 & & 24602 & 7263 \\ 
& XIS1 & 5x5 & & 3648 & 1526 \\ 
& XIS2 & 3x3 & & 70406 & 33157 \\ 
& XIS2 & 5x5 & & 24602 & 7263 \\ 
& XIS2 & 5x5 & & 3648 & 1526 \\ 
& XIS3 & 3x3 & & 68264 & 31279 \\ 
& XIS3 & 3x3 & & 1846 & 1846 \\ 
& XIS3 & 5x5 & & 24602 & 7263 \\ 
& XIS3 & 5x5 & & 3648 & 1526 \\ 
\hline
\end{tabular}
}
\subfloat{
\begin{tabular}[t]{|l l l l l l|}
\hline
Obs ID & Instr & Mode & Date start & Telapse & Ontime \\ 
\hline 
101005070 & XIS0 & 3x3 & 2006-10-21T15:36:54 & 15064 & 12456 \\ 
& XIS0 & 3x3 & & 24792 & 18680 \\ 
& XIS0 & 5x5 & & 7674 & 6034 \\ 
& XIS1 & 3x3 & & 15064 & 12440 \\ 
& XIS1 & 3x3 & & 24792 & 18656 \\ 
& XIS1 & 5x5 & & 7674 & 6034 \\ 
& XIS2 & 3x3 & & 15064 & 12456 \\ 
& XIS2 & 3x3 & & 24792 & 18680 \\ 
& XIS2 & 5x5 & & 7674 & 6034 \\ 
& XIS3 & 3x3 & & 15064 & 12456 \\ 
& XIS3 & 3x3 & & 24784 & 18672 \\ 
& XIS3 & 5x5 & & 7674 & 6034 \\ 
101005090 & XIS0 & 3x3 & 2006-12-13T18:53:16 & 29180 & 28226 \\ 
& XIS1 & 3x3 & & 29180 & 28226 \\ 
& XIS3 & 3x3 & & 29180 & 28226 \\ 
101005100 & XIS0 & 3x3 & 2007-01-15T03:20:53 & 44089 & 22614 \\ 
& XIS1 & 3x3 & & 44089 & 22614 \\ 
& XIS3 & 3x3 & & 44089 & 22614 \\ 
101005110 & XIS0 & 3x3 & 2007-02-10T22:13:47 & 75499 & 36102 \\ 
& XIS1 & 3x3 & & 75491 & 36094 \\ 
& XIS3 & 3x3 & & 75499 & 36102 \\ 
101005120 & XIS0 & 3x3 & 2007-03-18T21:11:19 & 31388 & 18242 \\ 
& XIS1 & 3x3 & & 31396 & 18250 \\ 
& XIS3 & 3x3 & & 31404 & 18258 \\ 
102001010 & XIS0 & 3x3 & 2007-04-10T10:35:08 & 30202 & 18115 \\ 
& XIS1 & 3x3 & & 30202 & 18115 \\ 
& XIS3 & 3x3 & & 30202 & 18115 \\ 
102002010 & XIS0 & 3x3 & 2007-06-13T10:10:12 & 60202 & 19155 \\ 
& XIS0 & 5x5 & & 25128 & 8717 \\ 
& XIS1 & 3x3 & & 60202 & 19155 \\ 
& XIS1 & 5x5 & & 25128 & 8717 \\ 
& XIS3 & 3x3 & & 60202 & 19155 \\ 
& XIS3 & 5x5 & & 25128 & 8717 \\ 
102003010 & XIS0 & 3x3 & 2007-08-12T05:21:09 & 80306 & 39490 \\ 
& XIS1 & 3x3 & & 80298 & 39482 \\ 
& XIS3 & 3x3 & & 80306 & 39490 \\ 
102004010 & XIS0 & 3x3 & 2007-10-25T12:24:45 & 70817 & 20679 \\ 
& XIS0 & 5x5 & & 18945 & 5496 \\ 
& XIS1 & 3x3 & & 70817 & 20679 \\ 
& XIS1 & 5x5 & & 18945 & 5496 \\ 
& XIS3 & 3x3 & & 70817 & 20679 \\ 
& XIS3 & 5x5 & & 18945 & 5496 \\ 
102005010 & XIS0 & 3x3 & 2007-12-01T19:25:40 & 51709 & 24767 \\ 
& XIS1 & 3x3 & & 51717 & 24791 \\ 
& XIS3 & 3x3 & & 51725 & 24791 \\ 
102006010 & XIS0 & 3x3 & 2008-03-15T05:43:27 & 48686 & 26801 \\ 
& XIS0 & 5x5 & & 3246 & 1438 \\ 
& XIS1 & 3x3 & & 48686 & 26801 \\ 
& XIS1 & 5x5 & & 3246 & 1438 \\ 
& XIS3 & 3x3 & & 48686 & 26801 \\ 
& XIS3 & 5x5 & & 3246 & 1438 \\ 
102021010 & XIS0 & 3x3 & 2007-09-28T06:13:48 & 32517 & 24915 \\ 
& XIS1 & 3x3 & & 32517 & 24915 \\ 
& XIS3 & 3x3 & & 32517 & 24915 \\ 
102022010 & XIS0 & 3x3 & 2008-02-14T16:57:28 & 37839 & 20970 \\ 
& XIS0 & 5x5 & & 18148 & 5509 \\ 
& XIS1 & 3x3 & & 100646 & 36897 \\ 
& XIS1 & 5x5 & & 104399 & 11541 \\ 
& XIS3 & 3x3 & & 37839 & 20970 \\ 
& XIS3 & 5x5 & & 18148 & 5509 \\ 
103001010 & XIS0 & 3x3 & 2008-04-08T14:33:12 & 37825 & 18635 \\ 
& XIS0 & 5x5 & & 12486 & 3743 \\ 
& XIS1 & 3x3 & & 37809 & 18619 \\ 
& XIS1 & 5x5 & & 12486 & 3743 \\ 
& XIS3 & 3x3 & & 37825 & 18635 \\ 
& XIS3 & 5x5 & & 12486 & 3743 \\ 
103001020 & XIS0 & 3x3 & 2008-06-05T03:50:53 & 58075 & 14348 \\ 
& XIS0 & 5x5 & & 23267 & 6981 \\ 
& XIS1 & 3x3 & & 58059 & 14332 \\ 
& XIS1 & 5x5 & & 23267 & 6981 \\ 
& XIS3 & 3x3 & & 58075 & 14348 \\ 
& XIS3 & 5x5 & & 23267 & 6981 \\ 
103001030 & XIS0 & 3x3 & 2008-08-12T22:21:28 & 56892 & 16876 \\ 
& XIS0 & 5x5 & & 17269 & 4434 \\ 
& XIS1 & 3x3 & & 56892 & 16876 \\ 
& XIS1 & 5x5 & & 17269 & 4434 \\ 
& XIS3 & 3x3 & & 56884 & 16868 \\ 
& XIS3 & 5x5 & & 17269 & 4434 \\ 
\hline
\end{tabular}
}
\end{scriptsize}
\label{tab:suz_obs}
\end{table}
\begin{table}
\caption{Table~A.1: (continued)}
\begin{scriptsize}
\subfloat{
\begin{tabular}[t]{|l l l l l l|}
\hline
Obs ID & Instr & Mode & Date start & Telapse & Ontime \\ 
\hline 
103001040 & XIS0 & 3x3 & 2008-10-22T02:31:56 & 41072 & 24381 \\ 
& XIS0 & 5x5 & & 1024 & 1024 \\ 
& XIS1 & 3x3 & & 41072 & 24381 \\ 
& XIS1 & 5x5 & & 1024 & 1024 \\ 
& XIS3 & 3x3 & & 41072 & 24381 \\ 
& XIS3 & 5x5 & & 1024 & 1024 \\ 
103001050 & XIS0 & 3x3 & 2008-12-13T13:33:21 & 41409 & 23416 \\ 
& XIS0 & 5x5 & & 18211 & 6218 \\ 
& XIS1 & 3x3 & & 41409 & 23416 \\ 
& XIS1 & 5x5 & & 18211 & 6218 \\ 
& XIS3 & 3x3 & & 41409 & 23400 \\ 
& XIS3 & 5x5 & & 18211 & 6218 \\ 
103001060 & XIS0 & 3x3 & 2009-03-09T03:07:54 & 42302 & 23860 \\ 
& XIS1 & 3x3 & & 42302 & 23844 \\ 
& XIS3 & 3x3 & & 42302 & 23868 \\ 
104005010 & XIS0 & 3x3 & 2009-04-23T15:17:13 & 42736 & 12225 \\ 
& XIS0 & 5x5 & & 19096 & 7215 \\ 
& XIS1 & 3x3 & & 82529 & 37813 \\ 
& XIS1 & 5x5 & & 19096 & 7215 \\ 
& XIS3 & 3x3 & & 42736 & 12225 \\ 
& XIS3 & 5x5 & & 19096 & 7215 \\ 
104006010 & XIS0 & 3x3 & 2009-06-26T03:42:21 & 32790 & 21083 \\ 
& XIS1 & 3x3 & & 32790 & 21083 \\ 
& XIS3 & 3x3 & & 32790 & 21083 \\ 
104007010 & XIS0 & 3x3 & 2009-10-26T18:14:08 & 56690 & 16256 \\ 
& XIS0 & 5x5 & & 13067 & 4127 \\ 
& XIS1 & 3x3 & & 56690 & 16256 \\ 
& XIS1 & 5x5 & & 13067 & 4127 \\ 
& XIS3 & 3x3 & & 56682 & 16248 \\ 
& XIS3 & 5x5 & & 13067 & 4127 \\ 
104008010 & XIS0 & 3x3 & 2009-10-11T14:30:55 & 35836 & 24727 \\ 
& XIS1 & 3x3 & & 67352 & 41718 \\ 
& XIS1 & 5x5 & & 37449 & 7923 \\ 
& XIS3 & 3x3 & & 35836 & 24727 \\ 
104009010 & XIS0 & 3x3 & 2009-12-25T21:59:01 & 35015 & 18297 \\ 
& XIS0 & 5x5 & & 12010 & 3696 \\ 
& XIS1 & 3x3 & & 35007 & 18289 \\ 
& XIS1 & 5x5 & & 12010 & 3696 \\ 
& XIS3 & 3x3 & & 35015 & 18297 \\ 
& XIS3 & 5x5 & & 12010 & 3696 \\ 
104010010 & XIS0 & 3x3 & 2010-02-04T17:50:39 & 56815 & 14909 \\ 
& XIS0 & 5x5 & & 17561 & 5600 \\ 
& XIS1 & 3x3 & & 56807 & 14901 \\ 
& XIS1 & 5x5 & & 17561 & 5600 \\ 
& XIS3 & 3x3 & & 56815 & 14909 \\ 
& XIS3 & 5x5 & & 17561 & 5600 \\ 
104011010 & XIS0 & 3x3 & 2009-10-20T19:14:20 & 31735 & 14622 \\ 
& XIS0 & 5x5 & & 12664 & 4416 \\ 
& XIS1 & 3x3 & & 31735 & 15568 \\ 
& XIS1 & 5x5 & & 12664 & 4416 \\ 
& XIS3 & 3x3 & & 31735 & 15568 \\ 
& XIS3 & 5x5 & & 12664 & 4416 \\ 
104012010 & XIS0 & 3x3 & 2010-02-05T10:26:37 & 48368 & 23494 \\ 
& XIS1 & 3x3 & & 48376 & 23502 \\ 
& XIS3 & 3x3 & & 48366 & 23492 \\ 
104014010 & XIS1 & 3x3 & 2009-12-01T21:39:19 & 88538 & 51948 \\ 
& XIS1 & 5x5 & & 18663 & 10711 \\ 
105004010 & XIS0 & 3x3 & 2010-04-05T00:02:07 & 40761 & 21593 \\ 
& XIS1 & 3x3 & & 40761 & 21593 \\ 
& XIS3 & 3x3 & & 40753 & 21593 \\ 
105004020 & XIS0 & 3x3 & 2010-06-19T03:06:41 & 53519 & 13148 \\ 
& XIS0 & 5x5 & & 19007 & 6090 \\ 
& XIS1 & 3x3 & & 53519 & 13148 \\ 
& XIS1 & 5x5 & & 19007 & 6090 \\ 
& XIS3 & 3x3 & & 53511 & 13140 \\ 
& XIS3 & 5x5 & & 19007 & 6090 \\ 
105004040 & XIS0 & 3x3 & 2010-10-26T19:30:06 & 29894 & 16549 \\ 
& XIS0 & 5x5 & & 11831 & 4071 \\ 
& XIS1 & 3x3 & & 29902 & 15948 \\ 
& XIS1 & 5x5 & & 11831 & 4071 \\ 
& XIS3 & 3x3 & & 29910 & 16573 \\ 
& XIS3 & 5x5 & & 11831 & 4071 \\ 
105004050 & XIS0 & 3x3 & 2010-12-09T00:36:56 & 39870 & 20071 \\ 
& XIS1 & 3x3 & & 37253 & 19985 \\ 
& XIS3 & 3x3 & & 37237 & 19969 \\ 
105004060 & XIS0 & 3x3 & 2011-02-08T00:21:57 & 33863 & 17259 \\ 
& XIS1 & 3x3 & & 33847 & 17243 \\ 
& XIS3 & 3x3 & & 33863 & 17259 \\ 
\hline
\end{tabular}
}
\subfloat{
\begin{tabular}[t]{|l l l l l l|}
\hline
Obs ID & Instr & Mode & Date start & Telapse & Ontime \\ 
\hline 
105005010 & XIS0 & 3x3 & 2010-06-19T18:54:20 & 44296 & 20278 \\ 
& XIS1 & 3x3 & & 44296 & 20278 \\ 
& XIS3 & 3x3 & & 44296 & 20278 \\ 
105005020 & XIS0 & 3x3 & 2010-12-09T11:43:15 & 63265 & 16312 \\ 
& XIS0 & 5x5 & & 23142 & 6159 \\ 
& XIS1 & 3x3 & & 63265 & 16312 \\ 
& XIS1 & 5x5 & & 23142 & 6159 \\ 
& XIS3 & 3x3 & & 63265 & 16312 \\ 
& XIS3 & 5x5 & & 23142 & 6159 \\ 
105006020 & XIS0 & 3x3 & 2010-12-07T19:16:57 & 104320 & 40729 \\ 
& XIS1 & 3x3 & & 104320 & 40672 \\ 
105006030 & XIS0 & 3x3 & 2010-08-29T18:28:07 & 93168 & 36454 \\ 
& XIS1 & 3x3 & & 93152 & 36438 \\ 
106002010 & XIS0 & 3x3 & 2011-04-11T19:52:58 & 31879 & 17288 \\ 
& XIS0 & 5x5 & & 11504 & 3098 \\ 
& XIS1 & 3x3 & & 31879 & 17288 \\ 
& XIS1 & 5x5 & & 11504 & 3098 \\ 
& XIS3 & 3x3 & & 31879 & 17288 \\ 
& XIS3 & 5x5 & & 11504 & 3098 \\ 
106002020 & XIS0 & 3x3 & 2011-06-29T02:20:44 & 41528 & 28790 \\ 
& XIS1 & 3x3 & & 41536 & 28798 \\ 
& XIS3 & 3x3 & & 41520 & 28782 \\ 
106002030 & XIS0 & 3x3 & 2011-10-14T15:30:43 & 29488 & 26701 \\ 
& XIS0 & 5x5 & & 7796 & 6124 \\ 
& XIS1 & 3x3 & & 29480 & 26693 \\ 
& XIS1 & 5x5 & & 7796 & 6124 \\ 
& XIS3 & 3x3 & & 29488 & 26701 \\ 
& XIS3 & 5x5 & & 7796 & 6124 \\ 
106002040 & XIS0 & 3x3 & 2012-03-17T09:12:28 & 49361 & 27563 \\ 
& XIS0 & 5x5 & & 14922 & 4831 \\ 
& XIS1 & 3x3 & & 49361 & 27547 \\ 
& XIS1 & 5x5 & & 14922 & 4831 \\ 
& XIS3 & 3x3 & & 49361 & 27563 \\ 
& XIS3 & 5x5 & & 14922 & 4831 \\ 
106002050 & XIS0 & 3x3 & 2011-04-12T08:28:16 & 32188 & 17344 \\ 
& XIS0 & 5x5 & & 11690 & 3481 \\ 
& XIS1 & 3x3 & & 32188 & 17344 \\ 
& XIS1 & 5x5 & & 9203 & 3386 \\ 
& XIS3 & 3x3 & & 32188 & 17344 \\ 
& XIS3 & 5x5 & & 11679 & 3470 \\ 
106003010 & XIS0 & 3x3 & 2011-04-12T21:21:13 & 26192 & 12419 \\ 
& XIS0 & 5x5 & & 15076 & 4410 \\ 
& XIS1 & 3x3 & & 26192 & 13867 \\ 
& XIS1 & 5x5 & & 15076 & 4410 \\ 
& XIS3 & 3x3 & & 26192 & 13867 \\ 
& XIS3 & 5x5 & & 15076 & 4410 \\ 
106003020 & XIS0 & 3x3 & 2011-10-15T02:11:09 & 31064 & 26261 \\ 
& XIS0 & 5x5 & & 7640 & 5304 \\ 
& XIS3 & 3x3 & & 31064 & 26357 \\ 
& XIS3 & 5x5 & & 7648 & 5319 \\ 
107002010 & XIS0 & 3x3 & 2012-04-22T22:13:35 & 49254 & 26319 \\ 
& XIS0 & 5x5 & & 12787 & 4151 \\ 
& XIS1 & 3x3 & & 49254 & 26287 \\ 
& XIS1 & 5x5 & & 12787 & 4151 \\ 
& XIS3 & 3x3 & & 49254 & 26311 \\ 
& XIS3 & 5x5 & & 12787 & 4151 \\ 
107002020 & XIS0 & 3x3 & 2012-06-25T09:02:38 & 51935 & 26992 \\ 
& XIS0 & 5x5 & & 16223 & 3736 \\ 
& XIS1 & 3x3 & & 51935 & 26992 \\ 
& XIS1 & 5x5 & & 16222 & 3735 \\ 
& XIS3 & 3x3 & & 51935 & 26992 \\ 
& XIS3 & 5x5 & & 16223 & 3736 \\ 
107002030 & XIS0 & 3x3 & 2012-10-29T02:53:50 & 51192 & 26841 \\ 
& XIS0 & 5x5 & & 17724 & 5269 \\ 
& XIS1 & 3x3 & & 51190 & 26839 \\ 
& XIS1 & 5x5 & & 17724 & 5269 \\ 
& XIS3 & 3x3 & & 51206 & 26834 \\ 
& XIS3 & 5x5 & & 17724 & 5269 \\ 
107002040 & XIS0 & 3x3 & 2013-03-13T21:17:23 & 64330 & 26441 \\ 
& XIS0 & 5x5 & & 17450 & 6160 \\ 
& XIS1 & 3x3 & & 64330 & 26449 \\ 
& XIS1 & 5x5 & & 17450 & 6160 \\ 
& XIS3 & 3x3 & & 64330 & 26449 \\ 
& XIS3 & 5x5 & & 17450 & 6160 \\ 
107003010 & XIS0 & 3x3 & 2012-04-23T17:30:23 & 60725 & 30367 \\ 
& XIS0 & 5x5 & & 4160 & 1761 \\ 
& XIS1 & 3x3 & & 60725 & 30367 \\ 
& XIS1 & 5x5 & & 4160 & 1761 \\ 
& XIS3 & 3x3 & & 60725 & 30367 \\ 
& XIS3 & 5x5 & & 4160 & 1761 \\ 
\hline
\end{tabular}
}
\end{scriptsize}
\label{tab:xmm_obs}
\end{table}

\begin{table}
\caption{}
\begin{scriptsize}
\begin{minipage}[t]{0.47\linewidth}\flushleft
\caption{Table~A.1: (continued)\newline}
\begin{tabular}[t]{|l l l l l l|}
\hline
Obs ID & Instr & Mode & Date start & Telapse & Ontime \\ 
\hline 
107003020 & XIS0 & 3x3 & 2012-10-29T22:40:18 & 54329 & 29143 \\ 
& XIS0 & 5x5 & & 60336 & 2108 \\ 
& XIS1 & 3x3 & & 54329 & 29143 \\ 
& XIS1 & 5x5 & & 60336 & 2108 \\ 
& XIS3 & 3x3 & & 54329 & 29143 \\ 
& XIS3 & 5x5 & & 60336 & 2108 \\ 
108002010 & XIS0 & 3x3 & 2013-04-27T11:03:35 & 66770 & 29345 \\ 
& XIS1 & 3x3 & & 64426 & 29323 \\ 
& XIS3 & 3x3 & & 66770 & 29337 \\ 
108002020 & XIS0 & 3x3 & 2013-06-26T14:09:57 & 65655 & 23429 \\ 
& XIS0 & 5x5 & & 23498 & 9715 \\ 
& XIS1 & 3x3 & & 65635 & 23427 \\ 
& XIS1 & 5x5 & & 23498 & 9715 \\ 
& XIS3 & 3x3 & & 65632 & 23424 \\ 
& XIS3 & 5x5 & & 23498 & 9715 \\ 
108002030 & XIS0 & 3x3 & 2013-09-28T14:05:19 & 76513 & 25446 \\ 
& XIS0 & 5x5 & & 23246 & 6513 \\ 
& XIS1 & 3x3 & & 76513 & 25446 \\ 
& XIS1 & 5x5 & & 23246 & 6513 \\ 
& XIS3 & 3x3 & & 76551 & 25448 \\ 
& XIS3 & 5x5 & & 23246 & 6513 \\ 
108002040 & XIS0 & 3x3 & 2014-03-15T15:45:37 & 59729 & 29279 \\ 
& XIS1 & 3x3 & & 59737 & 29279 \\ 
& XIS3 & 3x3 & & 59737 & 29287 \\ 
108003010 & XIS0 & 3x3 & 2013-04-28T05:38:23 & 65824 & 27443 \\ 
& XIS1 & 3x3 & & 65829 & 30785 \\ 
& XIS3 & 3x3 & & 65829 & 30785 \\ 
108003020 & XIS0 & 3x3 & 2013-09-29T11:39:14 & 77786 & 33613 \\ 
& XIS1 & 3x3 & & 77786 & 33613 \\ 
& XIS3 & 3x3 & & 77786 & 33613 \\ 
109001010 & XIS0 & 3x3 & 2014-04-21T05:39:51 & 74497 & 24142 \\ 
& XIS0 & 5x5 & & 23019 & 6596 \\ 
& XIS1 & 3x3 & & 74497 & 24150 \\ 
& XIS1 & 5x5 & & 23019 & 6596 \\ 
& XIS3 & 3x3 & & 74497 & 24150 \\ 
& XIS3 & 5x5 & & 23019 & 6596 \\ 
109001030 & XIS0 & 3x3 & 2015-04-03T16:10:16 & 57180 & 27957 \\ 
& XIS0 & 5x5 & & 5838 & 1773 \\ 
& XIS1 & 3x3 & & 57180 & 27957 \\ 
& XIS1 & 5x5 & & 5838 & 1773 \\ 
& XIS3 & 3x3 & & 57180 & 27957 \\ 
& XIS3 & 5x5 & & 5838 & 1773 \\ 
109002010 & XIS0 & 3x3 & 2014-04-22T03:19:30 & 75008 & 24998 \\ 
& XIS0 & 5x5 & & 23039 & 6686 \\ 
& XIS1 & 3x3 & & 74994 & 24984 \\ 
& XIS1 & 5x5 & & 23039 & 6686 \\ 
& XIS3 & 3x3 & & 75008 & 24998 \\ 
& XIS3 & 5x5 & & 23039 & 6686 \\ 
109002020 & XIS0 & 3x3 & 2014-10-29T05:35:13 & 71936 & 17151 \\ 
& XIS0 & 5x5 & & 23263 & 6684 \\ 
& XIS1 & 3x3 & & 71938 & 24660 \\ 
& XIS1 & 5x5 & & 23263 & 6684 \\ 
& XIS3 & 3x3 & & 71938 & 24660 \\ 
& XIS3 & 5x5 & & 23263 & 6684 \\ 
109003010 & XIS0 & 3x3 & 2015-04-07T13:56:45 & 66049 & 31388 \\ 
& XIS0 & 5x5 & & 6299 & 2001 \\ 
& XIS1 & 3x3 & & 66049 & 31388 \\ 
& XIS1 & 5x5 & & 6307 & 2009 \\ 
\hline 

\end{tabular}
\end{minipage}
\begin{minipage}[t]{0.47\linewidth}\flushleft
\caption{Table~A.2: Summary of the XMM-Newton observations used in this work (see text for more details).}
\begin{tabular}[t]{|l l l l l |}
\hline
Obs ID & ExpID & Date start & Ontime & Filter\\ 
\hline 
0123110201 & PNS001 & 2000-04-16T19:55:28.094 & 19300  & Thin1 \\ 
& M1S003 & 2000-04-16T20:08:21.126 & 18700 & Thin1 \\ 
& M2S005 & 2000-04-16T20:08:22.126 & 18700 & Thin1 \\ 
0123110301 & PNS002 & 2000-04-17T04:29:41.335 & 17240  & Medium \\ 
& M1S004 & 2000-04-17T04:42:34.365 & 16637 & Medium \\ 
& M2S006 & 2000-04-17T04:42:33.365 & 16645 & Medium \\ 
0135720601 & PNS001 & 2001-04-15T01:19:17.869 & 16004  & Thin1 \\ 
& PNS009 & 2001-04-14T21:02:04.263 & 12497 & Thin1 \\ 
& M1S003 & 2001-04-14T20:46:14.225 & 32705 & Thin1 \\ 
& M2S005 & 2001-04-14T20:46:14.225 & 32705 & Thin1 \\ 
0135720801 & PNS001 & 2001-12-25T18:57:56.487 & 31265  & Thin1 \\ 
& M1S003 & 2001-12-25T18:42:00.509 & 32064 & Thin1 \\ 
& M2S005 & 2001-12-25T18:42:00.509 & 32064 & Thin1 \\ 
0135720901 & PNS001 & 2002-04-20T23:00:53.809 & 12589  & Thin1 \\ 
& M2S005 & 2002-04-20T22:27:31.728 & 12680 & Thin1 \\ 
0135721001 & PNS001 & 2002-05-18T11:43:55.750 & 12026  & Thin1 \\ 
& PNS017 & 2002-05-18T15:56:39.404 & 13000 & Thin1 \\ 
& M1S018 & 2002-05-18T15:02:24.266 & 16600 & Thin1 \\ 
& M2S005 & 2002-05-18T10:40:34.580 & 14758 & Thin1 \\ 
0135721101 & PNS001 & 2002-10-13T03:25:52.041 & 10199  & Thin1 \\ 
& PNS017 & 2002-10-13T06:54:57.588 & 10200 & Thin1 \\ 
& M1S003 & 2002-10-13T03:21:08.051 & 23600 & Thin1 \\ 
& M2S005 & 2002-10-13T03:21:08.051 & 23600 & Thin1 \\ 
0135721301 & PNS001 & 2002-12-14T03:56:35.528 & 10900  & Thin1 \\ 
& PNS017 & 2002-12-14T07:37:18.089 & 14184 & Thin1 \\ 
& M1S003 & 2002-12-14T03:51:50.538 & 28277 & Thin1 \\ 
& M2S005 & 2002-12-14T03:51:50.538 & 28275 & Thin1 \\ 
0135721401 & PNS017 & 2003-04-20T19:20:34.709 & 18685  & Medium \\ 
& PNU002 & 2003-04-20T14:55:28.035 & 13599 & Medium \\ 
0135721501 & PNS001 & 2003-10-27T08:18:11.876 & 28292  & Thick \\ 
& M1S003 & 2003-10-27T07:55:47.931 & 29894 & Thin1 \\ 
& M2S005 & 2003-10-27T07:55:47.931 & 29883 & Thin1 \\ 
0135721701 & PNS001 & 2003-11-16T06:33:58.358 & 26449  & Thick \\ 
& M1S003 & 2003-11-16T06:11:31.414 & 27121 & Thin1 \\ 
& M2S005 & 2003-11-16T06:11:32.414 & 27124 & Thin1 \\ 
0135721901 & PNS001 & 2004-04-28T07:29:19.755 & 31643  & Thick \\ 
0135722001 & PNS001 & 2004-10-26T07:26:05.704 & 29517  & Thick \\ 
& M1S003 & 2004-10-26T06:57:21.774 & 31520 & Thin1 \\ 
& M2S005 & 2004-10-26T06:57:21.774 & 31520 & Thin1 \\ 
0135722101 & PNS001 & 2004-11-07T04:00:28.611 & 29837  & Thin1 \\ 
& M1S003 & 2004-11-07T03:38:04.665 & 31439 & Thin1 \\ 
& M2S005 & 2004-11-07T03:38:03.665 & 31428 & Thin1 \\ 
0135722201 & PNS001 & 2004-11-07T13:28:47.166 & 10466  & Thin2 \\ 
& PNU014 & 2004-11-07T16:36:10.671 & 5215 & Thin1 \\ 
& PNU027 & 2004-11-07T18:15:45.405 & 8298 & Thin2 \\ 
& M1U002 & 2004-11-07T13:09:48.216 & 22470 & UNDEF. \\ 
& M1U003 & 2004-11-07T21:15:26.919 & 2200 & UNDEF. \\ 
& M2S005 & 2004-11-07T13:06:22.225 & 22764 & Thin1 \\ 
& M2U002 & 2004-11-07T21:18:01.912 & 2100 & Thin1 \\ 
0135722301 & PNS001 & 2004-11-07T22:58:08.619 & 26842  & Thin1 \\ 
& M1S003 & 2004-11-07T22:35:43.681 & 31313 & Thin1 \\ 
& M2U002 & 2004-11-07T22:44:54.656 & 30761 & UNDEF. \\ 
0135722401 & PNS001 & 2004-10-14T09:10:52.339 & 30522  & Thick \\ 
& M1S003 & 2004-10-14T09:05:26.352 & 30719 & Thick \\ 
& M2S005 & 2004-10-14T09:05:26.352 & 30729 & Thick \\ 
0135722501 & PNS001 & 2005-04-17T22:43:29.224 & 34719  & Thin1 \\ 
0135722601 & PNS001 & 2005-11-05T06:50:36.089 & 29947  & Medium \\ 
& M1S003 & 2005-11-05T06:45:10.103 & 30146 & Thin1 \\ 
& M2S005 & 2005-11-05T06:45:11.103 & 30143 & Thin1 \\ 
0135722701 & PNS001 & 2006-04-20T02:29:36.236 & 30000  & Thin1 \\ 
& M1S003 & 2006-04-20T02:24:10.223 & 30200 & Thin1 \\ 
& M2S005 & 2006-04-20T02:24:09.223 & 30200 & Thin1 \\ 
0412980101 & PNS001 & 2006-11-05T01:00:54.354 & 31868  & Medium \\ 
& M1S002 & 2006-11-05T00:55:28.365 & 32065 & Thin1 \\ 
& M2S003 & 2006-11-05T00:55:28.365 & 32074 & Thin1 \\ 
0412980201 & PNS001 & 2007-04-25T12:41:15.764 & 35867  & Thin1 \\ 
& M1S002 & 2007-04-25T12:35:49.748 & 36062 & Thin1 \\ 
& M2S003 & 2007-04-25T12:35:49.748 & 36070 & Thin1 \\ 
0412980301 & PNS001 & 2007-10-26T09:55:18.821 & 36593  & Medium \\ 
& M1S002 & 2007-10-26T09:49:52.834 & 36795 & Thin1 \\ 
& M2S003 & 2007-10-26T09:49:52.834 & 36794 & Thin1 \\ 
\hline
\end{tabular}
\tablefoot{Columns 1 to 5 show the observation ID,  the exposure ID, the start of the observation,  the exposure time (in seconds), taken from the \texttt{ONTIME} keyword, and the filter used.}
\end{minipage}
\end{scriptsize}
\label{tab:chan_obs}
\end{table}

\begin{table}
\caption{}
\begin{scriptsize}
\begin{minipage}[t]{0.47\linewidth}\flushleft
\caption{Table~A.2: (continued)\newline}
\begin{tabular}[t]{|l l l l l |}
\hline
Obs ID & ExpID & Date start & Ontime & Mode\\ 
\hline 
0412980501 & PNS001 & 2008-04-19T09:27:21.525 & 29367  & Thin1 \\ 
& M1S002 & 2008-04-19T09:21:54.513 & 29565 & Thin1 \\ 
& M2S003 & 2008-04-19T09:21:54.513 & 29559 & Thin1 \\ 
0412980701 & M1S002 & 2008-11-14T19:48:41.122 & 28569  & Thin1 \\ 
& M2S003 & 2008-11-14T19:48:41.122 & 28573 & Thin1 \\ 
0412980801 & PNS001 & 2009-04-13T00:08:49.731 & 28398  & Thin1 \\ 
& M1S002 & 2009-04-13T00:03:23.718 & 28600 & Thin1 \\ 
& M2S003 & 2009-04-13T00:03:23.718 & 28600 & Thin1 \\ 
0412980901 & M1S002 & 2009-10-21T09:04:02.437 & 28600  & Thin1 \\ 
& M2S003 & 2009-10-21T09:04:02.437 & 28600 & Thin1 \\ 
0412981001 & PNS001 & 2010-04-21T01:41:11.495 & 30000  & Thin1 \\ 
& M1S002 & 2010-04-21T01:35:45.482 & 30200 & Thin1 \\ 
& M2S003 & 2010-04-21T01:35:45.482 & 30200 & Thin1 \\ 
0412981301 & M1S002 & 2010-10-18T21:42:28.952 & 7048  & Thin1 \\ 
& M1U002 & 2010-10-19T00:22:52.640 & 21994 & Thin1 \\ 
& M2S003 & 2010-10-18T21:42:28.952 & 7079 & Thin1 \\ 
& M2U002 & 2010-10-19T00:22:59.640 & 21995 & Thin1 \\ 
0412981401 & PNS001 & 2011-04-20T23:45:51.674 & 34600  & Thin1 \\ 
& M1S002 & 2011-04-20T23:40:25.660 & 34800 & Thin1 \\ 
& M2S003 & 2011-04-20T23:40:25.660 & 34800 & Thin1 \\ 
0412981501 & M1S002 & 2011-11-04T09:18:50.644 & 29900  & Thin1 \\ 
& M2S003 & 2011-11-04T09:18:50.644 & 29900 & Thin1 \\ 
0412981601 & M2U002 & 2011-11-04T18:40:40.206 & 28700  & Medium \\ 
0412981701 & PNS001 & 2012-12-06T16:27:27.059 & 14899  & Thin1 \\ 
& PNS012 & 2012-12-06T21:01:24.564 & 16685 & Medium \\ 
& PNS013 & 2012-12-07T02:05:21.973 & 20883 & Thick \\ 
& M1S002 & 2012-12-06T16:22:06.068 & 15200 & Thin1 \\ 
& M1S014 & 2012-12-06T20:47:06.591 & 16981 & Medium \\ 
& M1S015 & 2012-12-07T01:42:05.019 & 24380 & Thick \\ 
& M2S003 & 2012-12-06T16:22:06.068 & 15200 & Thin1 \\ 
& M2S016 & 2012-12-06T20:47:05.591 & 16983 & Medium \\ 
& M2S017 & 2012-12-07T01:42:06.019 & 11500 & Thick \\ 
0412982101 & PNS001 & 2013-11-07T04:46:07.490 & 31694  & Thin1 \\ 
& M1S002 & 2013-11-07T04:40:18.504 & 31900 & Thin1 \\ 
& M2S003 & 2013-11-07T04:40:47.502 & 31888 & Thin1 \\ 
0412982201 & PNS001 & 2014-10-20T08:12:59.834 & 33390  & Medium \\ 
& M1S002 & 2014-10-20T08:07:09.848 & 33587 & Thin1 \\ 
& M2S003 & 2014-10-20T08:07:39.847 & 33589 & Thin1 \\ 
0412982301 & M1S002 & 2014-10-20T18:10:28.279 & 43479  & Thin1 \\ 
& M2S003 & 2014-10-20T18:10:58.278 & 43477 & Thin1 \\ 
0412982401 & PNS001 & 2015-10-30T17:01:28.666 & 37110  & Medium \\ 
& M1S002 & 2015-10-30T16:55:38.682 & 37310 & Thin1 \\ 
& M2S003 & 2015-10-30T16:56:08.680 & 37282 & Thin1 \\ 
0412982501 & M1S002 & 2015-10-28T18:08:45.639 & 33482  & Thin1 \\ 
& M2S003 & 2015-10-28T18:09:14.638 & 33483 & Thin1 \\ 
0412983201 & PNS001 & 2016-10-26T23:24:23.019 & 35768  & Medium \\ 
& M1S002 & 2016-10-26T23:18:53.031 & 35964 & Thin1 \\ 
& M2S003 & 2016-10-26T23:19:22.030 & 35962 & Thin1 \\ 
0412983301 & M1S002 & 2016-12-03T18:37:33.517 & 33584  & Thin1 \\ 
& M2S003 & 2016-12-03T18:38:02.516 & 33588 & Thin1 \\ 
0791580701 & PNS001 & 2016-04-26T04:41:47.236 & 30397  & Thick \\ 
0791580801 & PNS001 & 2016-04-26T13:35:09.695 & 12399  & Medium \\ 
0791580901 & PNS001 & 2016-04-26T17:28:31.303 & 12400  & Thin1 \\ 
0791581001 & PNS001 & 2016-04-26T21:21:51.894 & 30400  & Thick \\ 
0791581101 & PNS001 & 2016-04-27T06:15:12.184 & 14088  & Medium \\ 
0791581201 & PNS001 & 2016-04-27T10:36:52.779 & 17400  & Thin1 \\ 
\hline 

\end{tabular}
\end{minipage}
\begin{minipage}[t]{0.49\linewidth}\flushleft
\caption{Table~A.3: Summary of the Chandra observations used in this work (see text  for more details).}
\begin{tabular}[t]{|l l l l l l|}
\hline
Set & Obs ID & DetNam & Date start & Ontime & DataMode\\ 
\hline 
1 & 138 & ACIS-235678 & 1999-08-23T16:04:07 & 9757 & FAINT\\ 
& 1231 & ACIS-235678 & 1999-08-23T19:18:41 & 9760 & FAINT\\ 
2 & 134 & ACIS-012367 & 1999-08-27T10:51:54 & 9754 & FAINT\\ 
& 1234 & ACIS-012367 & 1999-08-27T13:59:32 & 9759 & FAINT\\ 
3 & 120 & ACIS-456789 & 1999-09-28T22:35:15 & 88208 & FAINT\\ 
4 & 968 & ACIS-456789 & 1999-10-08T20:58:01 & 49035 & FAINT\\ 
5 & 1423 & ACIS-56789 & 1999-11-01T16:09:36 & 19162 & FAINT\\ 
6 & 48 & ACIS-01236 & 2000-01-17T18:34:47 & 8678 & FAINT\\ 
& 49 & ACIS-01236 & 2000-01-17T21:31:53 & 9270 & FAINT\\ 
7 & 420 & ACIS-012367 & 2000-03-14T02:32:29 & 10396 & VFAINT\\ 
8 & 136 & ACIS-012367 & 2000-03-16T04:17:31 & 10051 & FAINT\\ 
9 & 140 & ACIS-012367 & 2000-04-04T05:52:37 & 8345 & FAINT\\ 
& 439 & ACIS-012367 & 2000-04-04T08:30:34 & 6954 & FAINT\\ 
& 440 & ACIS-012367 & 2000-04-04T10:43:54 & 6955 & FAINT\\ 
10 & 444 & ACIS-012367 & 2000-04-30T08:20:31 & 8479 & FAINT\\ 
& 445 & ACIS-012367 & 2000-04-30T11:01:42 & 8022 & FAINT\\ 
11 & 141 & ACIS-235678 & 2000-05-28T08:56:05 & 9833 & FAINT\\ 
& 1702 & ACIS-235678 & 2000-05-28T12:07:21 & 9632 & FAINT\\ 
12 & 1803 & ACIS-235678 & 2000-07-03T00:59:24 & 8083 & FAINT\\ 
& 1789 & ACIS-235678 & 2000-07-03T03:36:02 & 7674 & FAINT\\ 
& 1783 & ACIS-012367 & 2000-07-03T05:56:32 & 7869 & FAINT\\ 
& 1784 & ACIS-012367 & 2000-07-03T08:36:32 & 7674 & FAINT\\ 
& 1785 & ACIS-012367 & 2000-07-03T10:57:02 & 7680 & FAINT\\ 
& 1786 & ACIS-012367 & 2000-07-03T13:17:33 & 7680 & FAINT\\ 
& 1787 & ACIS-012367 & 2000-07-03T15:38:02 & 7690 & FAINT\\ 
13 & 1308 & ACIS-235678 & 2000-12-10T06:17:32 & 7955 & FAINT\\ 
& 1311 & ACIS-235678 & 2000-12-10T08:53:25 & 7754 & FAINT\\ 
& 1312 & ACIS-235678 & 2000-12-10T11:15:14 & 7757 & FAINT\\ 
& 1478 & ACIS-456789 & 2000-12-10T13:37:05 & 7948 & FAINT\\ 
& 1510 & ACIS-456789 & 2000-12-10T16:18:25 & 7754 & FAINT\\ 
& 1525 & ACIS-456789 & 2000-12-10T18:40:14 & 7757 & FAINT\\ 
& 1526 & ACIS-456789 & 2000-12-10T21:02:04 & 7757 & FAINT\\ 
14 & 1314 & ACIS-012367 & 2000-12-15T13:57:20 & 6954 & FAINT\\ 
& 1315 & ACIS-012367 & 2000-12-15T16:05:49 & 6957 & FAINT\\ 
& 1316 & ACIS-012367 & 2000-12-15T18:14:19 & 6957 & FAINT\\ 
& 1317 & ACIS-012367 & 2000-12-15T20:22:49 & 6957 & FAINT\\ 
& 1527 & ACIS-012367 & 2000-12-15T22:31:18 & 6960 & FAINT\\ 
& 1528 & ACIS-012367 & 2000-12-16T00:39:48 & 6960 & FAINT\\ 
& 1529 & ACIS-012367 & 2000-12-16T02:48:18 & 6960 & FAINT\\ 
15 & 1530 & ACIS-235678 & 2001-06-06T01:12:14 & 7704 & FAINT\\ 
& 1531 & ACIS-235678 & 2001-06-06T03:49:14 & 7507 & FAINT\\ 
& 1532 & ACIS-235678 & 2001-06-06T06:07:00 & 7514 & FAINT\\ 
& 1533 & ACIS-012367 & 2001-06-05T06:36:47 & 7510 & FAINT\\ 
& 1534 & ACIS-012367 & 2001-06-05T09:01:21 & 7514 & FAINT\\ 
& 1535 & ACIS-012367 & 2001-06-05T11:19:06 & 7504 & FAINT\\ 
& 1536 & ACIS-012367 & 2001-06-05T13:36:52 & 7510 & FAINT\\ 
& 1537 & ACIS-012367 & 2001-06-05T15:56:14 & 7513 & FAINT\\ 
& 1538 & ACIS-456789 & 2001-06-06T08:24:45 & 7706 & FAINT\\ 
& 1539 & ACIS-456789 & 2001-06-06T11:02:00 & 7510 & FAINT\\ 
& 1540 & ACIS-456789 & 2001-06-06T13:19:45 & 7514 & FAINT\\ 
& 1541 & ACIS-456789 & 2001-06-06T15:37:31 & 7514 & FAINT\\ 
& 1542 & ACIS-012367 & 2001-06-05T18:17:10 & 7514 & FAINT\\ 
& 1543 & ACIS-012367 & 2001-06-05T20:36:29 & 7514 & FAINT\\ 
& 1544 & ACIS-012367 & 2001-06-05T22:54:13 & 7510 & FAINT\\ 
16 & 2836 & ACIS-012367 & 2001-12-05T14:37:53 & 7552 & FAINT\\ 
& 2837 & ACIS-012367 & 2001-12-05T16:56:22 & 7558 & FAINT\\ 
& 2838 & ACIS-012367 & 2001-12-05T19:14:52 & 7558 & FAINT\\ 
& 2839 & ACIS-012367 & 2001-12-05T21:33:22 & 7558 & FAINT\\ 
& 2840 & ACIS-012367 & 2001-12-05T23:51:51 & 7558 & FAINT\\ 
& 2841 & ACIS-012367 & 2001-12-06T02:10:21 & 7558 & FAINT\\ 
& 2842 & ACIS-012367 & 2001-12-06T04:28:51 & 7558 & FAINT\\ 
& 2843 & ACIS-235678 & 2001-12-06T06:47:36 & 7760 & FAINT\\ 
& 2844 & ACIS-235678 & 2001-12-06T09:25:20 & 7552 & FAINT\\ 
& 2845 & ACIS-235678 & 2001-12-06T11:43:50 & 7558 & FAINT\\ 
17 & 2846 & ACIS-456789 & 2001-12-08T23:16:07 & 7949 & FAINT\\ 
& 2847 & ACIS-456789 & 2001-12-09T01:46:57 & 7552 & FAINT\\ 
& 2848 & ACIS-456789 & 2001-12-09T04:05:27 & 7558 & FAINT\\ 
& 2849 & ACIS-456789 & 2001-12-09T06:23:56 & 7302 & FAINT\\ 
18 & 2850 & ACIS-235678 & 2002-06-19T16:25:39 & 7859 & FAINT\\ 
& 2851 & ACIS-235678 & 2002-06-19T18:59:01 & 7654 & FAINT\\ 
& 2852 & ACIS-235678 & 2002-06-19T21:20:25 & 7661 & FAINT\\ 
\hline 

\end{tabular}
\tablefoot{Columns 3 to 5 list the time of the start of the observation, the exposure time (in seconds), taken from the \texttt{ONTIME} keyword, and the  Observing Mode (Faint and Very Faint format, where pixel values are read out from a 3x3, and 5x5 region surrounding the event, respectively).}
\end{minipage}
\end{scriptsize}
\end{table}

\begin{table}
\caption{Table~A.3: (continued)}
\begin{scriptsize}
\subfloat{
\begin{tabular}[t]{|l l l l l l|}
\hline
Set & Obs ID & DetNam & Date start & Ontime & DataMode\\ 
\hline 
19 & 2857 & ACIS-012367 & 2002-06-21T02:11:35 & 7651 & FAINT\\ 
& 2858 & ACIS-012367 & 2002-06-21T04:39:09 & 7654 & FAINT\\ 
& 2859 & ACIS-012367 & 2002-06-21T06:59:19 & 7658 & FAINT\\ 
& 2860 & ACIS-012367 & 2002-06-21T09:19:30 & 7654 & FAINT\\ 
& 2862 & ACIS-012367 & 2002-06-21T13:59:50 & 7658 & FAINT\\ 
& 2863 & ACIS-012367 & 2002-06-21T16:20:00 & 7658 & FAINT\\ 
& 2864 & ACIS-012367 & 2002-06-21T18:40:10 & 7658 & FAINT\\ 
20 & 2853 & ACIS-456789 & 2002-06-22T06:20:31 & 7856 & FAINT\\ 
& 2854 & ACIS-456789 & 2002-06-22T08:52:42 & 7654 & FAINT\\ 
& 2855 & ACIS-456789 & 2002-06-22T11:12:52 & 7658 & FAINT\\ 
& 2856 & ACIS-456789 & 2002-06-22T13:34:30 & 7660 & FAINT\\ 
21 & 3828 & ACIS-456789 & 2002-12-20T14:48:39 & 137658 & FAINT\\ 
22 & 3519 & ACIS-235678 & 2003-02-01T04:33:34 & 8112 & VFAINT\\ 
& 3520 & ACIS-235678 & 2003-02-01T07:06:41 & 7728 & VFAINT\\ 
& 3521 & ACIS-235678 & 2003-02-01T09:28:04 & 7731 & VFAINT\\ 
& 3523 & ACIS-235678 & 2003-02-01T11:49:27 & 7731 & VFAINT\\ 
& 3524 & ACIS-235678 & 2003-02-01T14:10:50 & 7728 & VFAINT\\ 
& 3522 & ACIS-456789 & 2003-02-01T16:32:13 & 7933 & VFAINT\\ 
& 3525 & ACIS-456789 & 2003-02-01T19:13:06 & 7728 & VFAINT\\ 
& 3535 & ACIS-012367 & 2003-02-01T21:34:29 & 7935 & VFAINT\\ 
& 3536 & ACIS-012367 & 2003-02-02T00:15:22 & 7725 & VFAINT\\ 
& 3537 & ACIS-012367 & 2003-02-02T02:36:45 & 7731 & VFAINT\\ 
& 3538 & ACIS-012367 & 2003-02-02T04:58:08 & 7731 & VFAINT\\ 
& 3539 & ACIS-012367 & 2003-02-02T07:19:31 & 7731 & VFAINT\\ 
& 3540 & ACIS-012367 & 2003-02-02T09:40:54 & 7731 & VFAINT\\ 
& 3541 & ACIS-012367 & 2003-02-02T12:02:17 & 7731 & VFAINT\\ 
& 3542 & ACIS-012367 & 2003-02-02T14:23:40 & 7731 & VFAINT\\ 
& 3543 & ACIS-012367 & 2003-02-02T16:45:03 & 7738 & VFAINT\\ 
23 & 3526 & ACIS-012367 & 2003-08-05T22:31:32 & 8181 & VFAINT\\ 
& 3527 & ACIS-012367 & 2003-08-06T01:03:11 & 7965 & VFAINT\\ 
24 & 3546 & ACIS-235678 & 2003-08-08T17:58:11 & 7962 & VFAINT\\ 
& 3547 & ACIS-235678 & 2003-08-08T20:39:32 & 7754 & VFAINT\\ 
& 3548 & ACIS-235678 & 2003-08-08T23:01:21 & 7754 & VFAINT\\ 
& 3528 & ACIS-012367 & 2003-08-09T01:23:12 & 7962 & VFAINT\\ 
& 3529 & ACIS-012367 & 2003-08-09T04:04:31 & 7750 & VFAINT\\ 
& 3530 & ACIS-012367 & 2003-08-09T06:26:21 & 7750 & VFAINT\\ 
25 & 3532 & ACIS-012367 & 2003-08-10T09:07:04 & 7750 & VFAINT\\ 
& 3533 & ACIS-012367 & 2003-08-10T11:28:53 & 7757 & VFAINT\\ 
& 3534 & ACIS-012367 & 2003-08-10T13:50:44 & 7757 & VFAINT\\ 
& 3544 & ACIS-235678 & 2003-08-10T16:12:49 & 7962 & VFAINT\\ 
26 & 5123 & ACIS-456789 & 2003-12-15T04:24:59 & 20582 & VFAINT\\ 
& 5124 & ACIS-456789 & 2003-12-15T10:30:54 & 8003 & VFAINT\\ 
& 5125 & ACIS-456789 & 2003-12-15T12:56:53 & 8010 & VFAINT\\ 
& 5126 & ACIS-456789 & 2003-12-15T15:22:54 & 8010 & VFAINT\\ 
& 5127 & ACIS-456789 & 2003-12-15T17:48:53 & 8010 & VFAINT\\ 
& 5128 & ACIS-456789 & 2003-12-15T20:14:53 & 8006 & VFAINT\\ 
& 5129 & ACIS-456789 & 2003-12-15T22:40:53 & 8010 & VFAINT\\ 
27 & 5153 & ACIS-012367 & 2003-12-16T22:24:48 & 7491 & VFAINT\\ 
& 5154 & ACIS-012367 & 2003-12-17T00:53:15 & 7546 & VFAINT\\ 
28 & 5147 & ACIS-012367 & 2003-12-19T01:55:58 & 8032 & VFAINT\\ 
& 5148 & ACIS-012367 & 2003-12-19T04:39:08 & 19529 & VFAINT\\ 
& 5149 & ACIS-012367 & 2003-12-19T10:17:19 & 7667 & VFAINT\\ 
& 5150 & ACIS-012367 & 2003-12-19T12:37:42 & 7670 & VFAINT\\ 
& 5151 & ACIS-012367 & 2003-12-19T14:58:05 & 7670 & VFAINT\\ 
29 & 5152 & ACIS-012367 & 2003-12-22T07:45:55 & 8080 & VFAINT\\ 
30 & 5251 & ACIS-456789 & 2003-12-24T20:43:02 & 7708 & VFAINT\\ 
& 5252 & ACIS-456789 & 2003-12-24T23:15:05 & 7549 & VFAINT\\ 
31 & 5131 & ACIS-456789 & 2004-04-05T04:47:50 & 8115 & VFAINT\\ 
32 & 5130 & ACIS-456789 & 2004-04-09T13:07:30 & 19661 & VFAINT\\ 
& 5132 & ACIS-456789 & 2004-04-09T18:55:24 & 7596 & VFAINT\\ 
& 5133 & ACIS-456789 & 2004-04-09T21:14:35 & 7597 & VFAINT\\ 
& 5134 & ACIS-456789 & 2004-04-09T23:33:45 & 7597 & VFAINT\\ 
33 & 5135 & ACIS-456789 & 2004-04-10T12:34:32 & 8246 & VFAINT\\ 
& 5136 & ACIS-456789 & 2004-04-10T15:35:17 & 8041 & VFAINT\\ 
& 5139 & ACIS-012367 & 2004-04-10T18:01:57 & 20682 & VFAINT\\ 
34 & 5143 & ACIS-012367 & 2004-04-26T10:49:08 & 7443 & VFAINT\\ 
& 5144 & ACIS-012367 & 2004-04-26T13:14:00 & 7235 & VFAINT\\ 
& 5145 & ACIS-012367 & 2004-04-26T15:27:11 & 7238 & VFAINT\\ 
35 & 5137 & ACIS-012367 & 2004-04-28T00:27:20 & 8141 & VFAINT\\ 
& 5138 & ACIS-012367 & 2004-04-28T03:02:20 & 7754 & VFAINT\\ 
36 & 5140 & ACIS-012367 & 2004-04-30T16:57:40 & 8144 & VFAINT\\ 
& 5141 & ACIS-012367 & 2004-04-30T19:32:39 & 7754 & VFAINT\\ 
& 5142 & ACIS-012367 & 2004-04-30T21:54:29 & 7757 & VFAINT\\ 
37 & 6050 & ACIS-012367 & 2004-12-13T03:35:35 & 7254 & VFAINT\\ 
38 & 6052 & ACIS-012367 & 2004-12-14T08:58:26 & 7642 & VFAINT\\ 
39 & 6074 & ACIS-456789 & 2004-12-16T18:02:39 & 20099 & VFAINT\\ 
& 6055 & ACIS-012367 & 2004-12-16T23:56:21 & 7955 & VFAINT\\ 
\hline 

\end{tabular}
}
\subfloat{
\begin{tabular}[t]{|l l l l l l|}
\hline
Set & Obs ID & DetNam & Date start & Ontime & DataMode\\ 
\hline 
40 & 6057 & ACIS-012367 & 2004-12-17T22:47:29 & 7955 & VFAINT\\ 
& 6075 & ACIS-456789 & 2004-12-18T01:21:24 & 7955 & VFAINT\\ 
41 & 6076 & ACIS-456789 & 2004-12-19T10:52:37 & 7952 & VFAINT\\ 
& 6077 & ACIS-456789 & 2004-12-19T13:26:12 & 7754 & VFAINT\\ 
& 6078 & ACIS-456789 & 2004-12-19T15:48:02 & 7757 & VFAINT\\ 
42 & 6079 & ACIS-456789 & 2004-12-20T08:06:36 & 7955 & VFAINT\\ 
& 6080 & ACIS-456789 & 2004-12-20T10:38:29 & 7753 & VFAINT\\ 
43 & 6051 & ACIS-012367 & 2005-01-12T04:25:55 & 18150 & VFAINT\\ 
& 6053 & ACIS-012367 & 2005-01-12T09:49:06 & 7254 & VFAINT\\ 
& 6054 & ACIS-012367 & 2005-01-12T12:02:36 & 7258 & VFAINT\\ 
44 & 6042 & ACIS-456789 & 2005-04-12T01:40:22 & 19146 & VFAINT\\ 
45 & 6043 & ACIS-456789 & 2005-04-18T08:43:21 & 7952 & VFAINT\\ 
46 & 6060 & ACIS-012367 & 2005-06-13T13:12:55 & 20058 & VFAINT\\ 
47 & 6750 & ACIS-012367 & 2006-03-14T08:41:34 & 7603 & VFAINT\\ 
& 6751 & ACIS-012367 & 2006-03-14T11:13:46 & 7254 & VFAINT\\ 
& 6752 & ACIS-012367 & 2006-03-14T13:27:16 & 7254 & VFAINT\\ 
& 6753 & ACIS-012367 & 2006-03-14T15:40:47 & 7258 & VFAINT\\ 
& 6754 & ACIS-012367 & 2006-03-14T17:54:17 & 7254 & VFAINT\\ 
& 6755 & ACIS-012367 & 2006-03-14T20:07:47 & 7254 & VFAINT\\ 
48 & 6758 & ACIS-456789 & 2006-03-19T04:28:05 & 8163 & VFAINT\\ 
& 6760 & ACIS-456789 & 2006-03-19T07:01:31 & 7753 & VFAINT\\ 
& 6761 & ACIS-456789 & 2006-03-19T09:23:21 & 7757 & VFAINT\\ 
& 6762 & ACIS-456789 & 2006-03-19T11:45:11 & 7757 & VFAINT\\ 
& 6763 & ACIS-456789 & 2006-03-19T14:07:01 & 7754 & VFAINT\\ 
& 6764 & ACIS-456789 & 2006-03-19T16:28:52 & 7757 & VFAINT\\ 
49 & 6759 & ACIS-456789 & 2006-03-21T23:28:29 & 18141 & VFAINT\\ 
50 & 6748 & ACIS-012367 & 2006-03-22T11:47:51 & 7951 & VFAINT\\ 
& 6749 & ACIS-012367 & 2006-03-22T14:25:07 & 19757 & VFAINT\\ 
51 & 6747 & ACIS-012367 & 2006-04-03T05:23:29 & 7376 & VFAINT\\ 
52 & 6757 & ACIS-012367 & 2006-06-05T13:19:43 & 20058 & VFAINT\\ 
53 & 6766 & ACIS-456789 & 2006-06-06T13:28:51 & 19949 & VFAINT\\ 
54 & 8361 & ACIS-012367 & 2007-02-05T02:20:53 & 20045 & VFAINT\\ 
55 & 8366 & ACIS-235678 & 2007-02-08T13:43:28 & 8342 & VFAINT\\ 
& 8367 & ACIS-235678 & 2007-02-08T16:22:01 & 7955 & VFAINT\\ 
56 & 8362 & ACIS-012367 & 2007-02-11T08:13:03 & 8957 & VFAINT\\ 
& 8363 & ACIS-012367 & 2007-02-11T11:00:38 & 8553 & VFAINT\\ 
& 8364 & ACIS-012367 & 2007-02-11T13:35:48 & 8557 & VFAINT\\ 
57 & 9691 & ACIS-01236 & 2008-02-04T23:02:12 & 8045 & VFAINT\\ 
& 9692 & ACIS-01236 & 2008-02-05T01:42:29 & 7774 & VFAINT\\ 
& 9693 & ACIS-01236 & 2008-02-05T04:04:18 & 7777 & VFAINT\\ 
58 & 9695 & ACIS-35678 & 2008-02-07T13:58:22 & 8032 & VFAINT\\ 
& 9696 & ACIS-35678 & 2008-02-07T16:38:39 & 7774 & VFAINT\\ 
59 & 10652 & ACIS-01236 & 2009-01-17T20:41:03 & 8175 & VFAINT\\ 
60 & 10650 & ACIS-01236 & 2009-02-16T18:17:32 & 8038 & VFAINT\\ 
61 & 10651 & ACIS-01236 & 2009-02-17T07:38:59 & 8043 & VFAINT\\ 
62 & 12147 & ACIS-456789 & 2011-02-11T17:01:11 & 150794 & FAINT\\ 
63 & 13411 & ACIS-5 & 2011-05-23T02:29:16 & 10028 & VFAINT\\ 
64 & 13409 & ACIS-5 & 2011-05-28T01:40:18 & 10028 & VFAINT\\ 
& 13410 & ACIS-5 & 2011-05-28T04:45:17 & 10028 & VFAINT\\ 
65 & 17687 & ACIS-3 & 2015-07-18T02:35:28 & 13966 & VFAINT\\ 
66 & 18419 & ACIS-7 & 2016-03-22T05:39:25 & 20063 & VFAINT\\ 
\hline 

\end{tabular}
}
\end{scriptsize}
\end{table}

\end{appendix}


\end{document}